# How might Driver Licensing and Vehicle Registration evolve if we adopt Autonomous Cars and Digital Identification?

Scott McLachlan

*Do not go gentle into that digital night,*
*Be more than your data shadow at the close of day,*
*Rage, rage against the dying of privacy.*[1]

> In 1982 the Australian House of Representatives Standing Committee on Road Safety stated: *"Licensing of drivers has traditionally been seen as a screening process, with the aim of keeping unsafe drivers off the road. However, at the present time no system of testing licence applicants has been developed that could be used to deny licences to those highly likely to be involved in accidents, without at the same time denying licences to very large numbers of drivers who would subsequently be accident free"*.

## 1. Introduction

In 1885 what is often described as the world's first *automobile*[2], a prototype horseless carriage propelled by a small internal combustion engine, was demonstrated to a group of onlookers in the city of Mannheim, Germany (Dietsche & Kuhlgatz, 2014). While others were also developing early prototype automobiles at that time, including now-famous names like Gottlieb Daimler and Wilhelm Maybach, it was Karl Benz's Patent Motorwagen that in January 1886 was to become the world's first produced and patented[3] automobile (Dietsche & Kuhlgatz, 2014; Schreiber, 2014). The Patent Motorwagen's initial public demonstration did not go as planned, and would leave Benz frustrated at its outward failure. The three-wheeled gasoline-powered vehicle was observed to produce copious clouds of malodourous smoke, and was noisy, slow and difficult to control (Ennajih et al, 2015). However, these issues became less significant after Benz, who on forgetting to steer during this maiden outing, crashed the vehicle into a wall (Duncan & Meals, 1995). Consequently, this resulted in another first for Benz. The world's first *automobile accident*. However, and more importantly for this work, these would not be the last time the term *world's first* would be associated with the Benz family name generally, and Karl Benz specifically.

---

[1] Adapted from *Do not go gentle into that good night* by Dylan Thomas.

[2] While there are descriptions and pictures of simple prototype steam and gasoline powered vehicles that existed prior to Benz's Patent Motorwagen, including Cugnot's 1769 steam-powered tricycle, Henry Taylor's 1867 Steam Buggy and Siegfried Marcus' 1860's benzene-powered motorcar, there was nothing to suggest these vehicles had been patented or produced for sale. The general consensus in the literature is that Benz's Patent Motorwagen was first in both regards.

[3] Patent DRP-37435: "automobile fuelled by gas" was issued on January 29, 1886.

After complaints about the noise and smell, and several accidents for which Benz was blamed, the Mayor of Mannheim banned the Patent Motorwagen from his city's main roads. In response, Benz requested and received the world's first *driving license* shown in Figure 1; a permit issued under authority of the Grand Duchy of Baden (Ennajih et al, 2015). However, and unlike today's driving licenses, this document was more about permitting the presence of Benz's Patent Motorwagen on Mannheim roads than providing certification for the human driver.

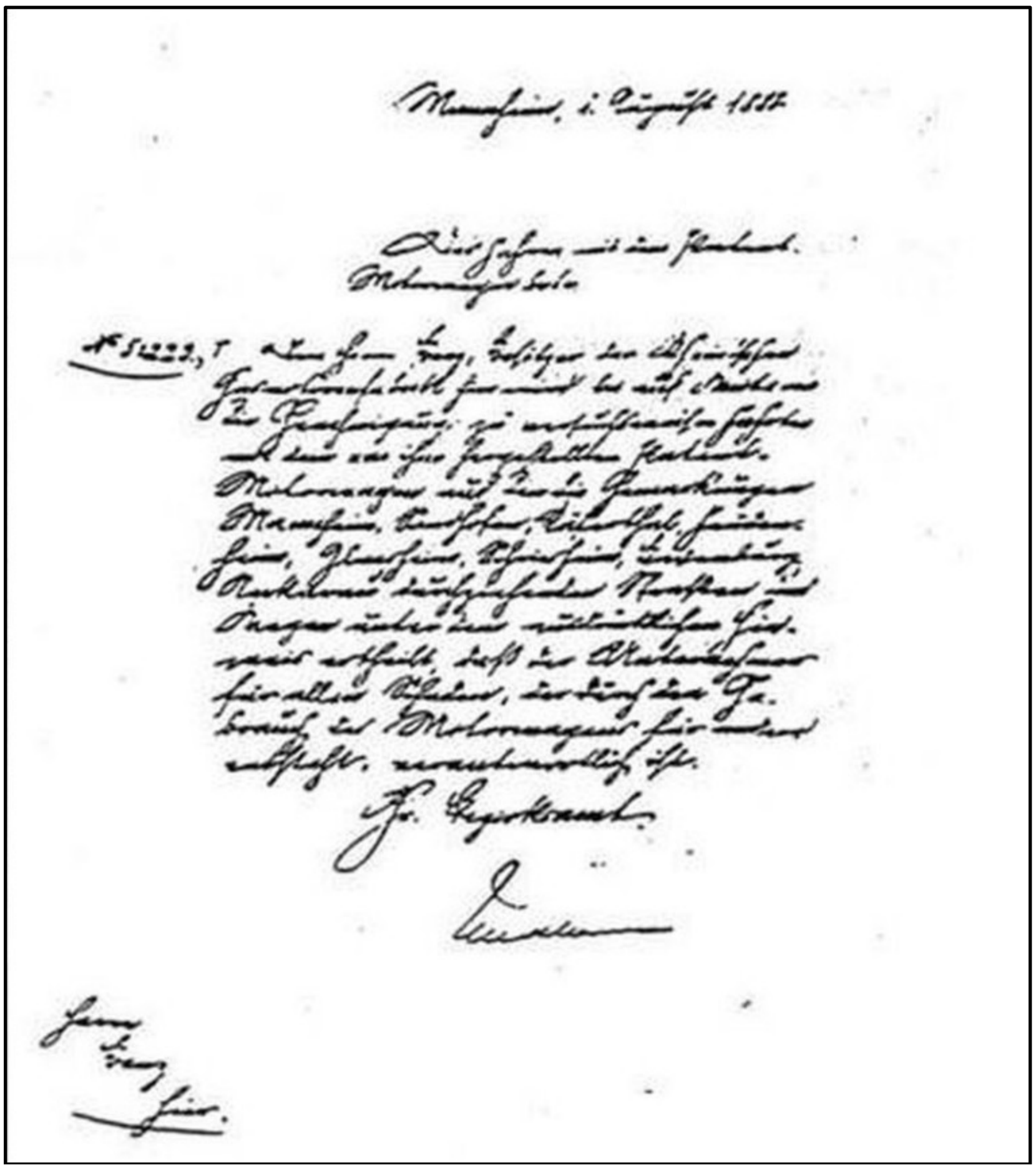

***Figure 1:*** *1888 Driving Permit issued to Karl Benz (source: Mercedes-Benz Museum)*

Even before their marriage in 1872 Bertha Benz (nee Ringer) had been a strong supporter of Benz's engineering ventures. In 1871 she used her dowry to buy out August Ritter, Benz's unreliable *Iron Foundry and Mechanical Workshop* business partner (Hunter, 2012; van der Kooij, 2021). This placed Benz in control of the firm and over the next decade he developed and registered successive inventions, and was awarded many patents, including for: (i) a speed regulation system; (ii) an ignition system using sparks with a battery; (iii) the spark plug; (iv) the carburettor; (v) the clutch; (vi) the gear shifter; and (vii) a water radiator (Oachs, 2019; Rajin, 2017). Even though Benz successfully demonstrated the third model of his Patent Motorwagen at the 1887 Paris Expo, true success only came after Bertha, along with Benz's two sons Eugen and Richard, made what is described by many as the world's first *long distance* *automobile trip* on August 5$^{th}$, 1888 (Hunter, 2012; Warsi et al, 2020). Her 194km round trip from Mannheim via Heidelberg to Florsheim and back demonstrated the feasibility of Benz's vehicle for travel, and the route is now signposted as the *Bertha Benz Memorial Route*. Popularity of the Patent Motorwagen grew, sales increased, and the motor vehicle and all its incumbent problems was well on the way to becoming the ubiquitous phenomenon it is today.

## 2. A Brief History of Vehicle Registration and the Driver's Licensing

Most early driver's licenses were not created to be identification documents and had little to do with safety, as they did not require specific training, knowledge or skills tests (Redman, 2008). They were primarily intended to signify to vehicle owners that driving should not be considered a right, but rather was a privilege: and the fees were certainly a welcome boost for government revenues[4] (Harper, 2006). Some variation of these formative vehicle registration and driving licensing requirements arose in countries all over the world, including:

**France:** When the *Paris Police Ordinance* came into effect on August 14, 1893 France possibly became the first country to mandate driving tests to receive the French *certificat de capacité de conduit* (*permis de conduire*) - effectively a competency certificate for driving (Carter, 2020; Rerat, 2021). Formal driving lessons were introduced in 1917 (Carter, 2020).

**Australia:** Australia is a federated collection of six states and ten territories - three internal and seven external. Queensland's *Brisbane Traffic Act* 1895 required drivers and conductors of public passenger and goods vehicles for hire to obtain a license from the Police Commissioners' Office, and from 1914 to submit to a driving test (Dukova, 2014). The first report of registration plates and a driver's license being issued to a private citizen was on September 10, 1906 in South Australia to Dr William Arthur Hargreaves[5]. Hargreaves, a chemist, had developed an early alternative fuel based on a mixture of molasses and petrol that he used to fuel his motor car. License plates and driver's licenses were introduced in Victoria and New South Wales in 1910[6]. However, while Queensland commenced issuing driver licenses in 1910 in response to the State's first recorded road fatality[7], *vehicle registrations* and *driver licensing* of private citizens was not mandated in law there until 1921 and 1922 respectively (Dukova, 2014). It was that same fatal motor vehicle accident in 1910 that led police the following year to commence driver testing, which, as stated above, was expanded three years later to include public passenger conveyance and vehicle for hire drivers[8].

**New Zealand (NZ):** The *Motor Car Regulation Act* 1902 required certification of drivers of public passenger and hire vehicles. Even though no legislation at the time imposed the requirement for a driving test for private citizens, from 1903 the Auckland Automobile Association issued a certificate of competency to those who passed their driving exam (Johnson, 2019). Local authorities (Councils) oversaw vehicle registration from 1905, and the *Motor Vehicle Act* 1924 introduced compulsory annual driver's licenses from 1925 (Pawson, 2014; Swarbrick, 2014). A centralised country-wide transport department was established and assumed both functions in 1929 (Pawson, 2014). Knowledge and practical driving skills tests became compulsory from June 1931[9]. In 1987 NZ became the first country in the world to adopt a *graduated driver licensing* system wherein novice and young drivers first receive training and experience under supervision from an experienced licensed driver such as a parent, before being given freedom, within constraints[10], to gain additional independent experience as a provisionally licensed driver (Waller, 2003).

---

[4] For example: amendments introduced to the legislation of Victoria, Australia in the *Motor Car Act* 1930 included 21 separate newly created or expanded fees or payments to be made by owners or vehicle operators into State revenues and 10 new or expanded penalties. Conversely, there were no amendments in the 1930 Act that increased vehicle safety or driver training and testing.

[5] https://twistedhistory.net.au/2017/09/10/the-first-australian-licence-plates-and-drivers-licence-issued-2/

[6] Ibid.

[7] https://www.qld.gov.au/transport/licensing/getting/about-queensland-driver-licence-cards

[8] Ibid.

**United States of America (USA):** Municipal governments initially took the lead in mandating registration and identification plates for automobiles and developing rules for driver licensing. One of the first was the City of Chicago in 1899 (Rao & Singh, 2001; Roots, 2005). On April 25th 1901 New York enacted the first State-wide regulation for vehicle registration (Marvin, 1987; Rao & Singh, 2001). In 1906 that New Jersey became the first to enact State-wide driver licensing regulations, which they followed in 1908 with a requirement for basic knowledge and driving skills tests (Rao & Singh, 2001). However, it wasn't until the 1930's that standardised driver education courses were developed[11].

**Canada:** The first automobiles arrived in Ontario in 1898 and the first provincial regulation governing use, registration and number plates came into effect in 1903[12]. While Ontario issued the first Canadian driver's licenses to chauffeurs from 1909, it was not until 1927 that private citizens were required to obtain a vehicle *operator's license*[13] and the first provincial *Highway Traffic Act* was passed. Beginner Permits were also introduced in 1927, which allowed new drivers to gain experience when supported and accompanied by a fully licensed driver. Two years later on June 12th, 1929 Alberta also commenced issuing driver's licenses printed on linen, with the first going to the then State Premier, Mr John Brownlee,[14].

**United Kingdom (UK):** While not explicitly stated in legislation, prior to 1896 the use of anything motorised except slow and heavy traction engines on UK highways was, in at least the most practical sense, prohibited[15] (Selby, 1906). Provisions introduced in *The Locomotives on Highways Act* 1896 (LOHA) exempted what it defined as *light locomotives*[16], and went on to class as *carriages*[17] in order to fall within that meaning in other UK laws and regulations, from almost all provisions of previous Acts relating to the use of locomotives on highways. The light locomotive is described as one *propelled by mechanical power* and LOHA's new

[9] https://www.nzta.govt.nz/roads-and-rail/research-and-data/fascinating-facts/regulations/

[10] These constraints include limitations on maximum vehicle power, speed and number of vehicle occupants, and a daylight hours curfew.

[11] https://www.motortrend.com/news/fast-facts-the-113-year-history-of-the-drivers-license-110875/

[12] http://www.mto.gov.on.ca/english/about/mto-100/index.shtml

[13] Ibid.

[14] https://calgaryherald.com/news/local-news/this-day-in-history-june-12-1929-albertas-first-drivers-licence-issued-on-linen

[15] Evidence of Mr T Wainwright, the first witness interviewed during conduct of the 1906 Royal Commission on Motor Cars, described:

"The law then in force with regard to the use of locomotives upon highways had been framed so as to permit of the use of road engines, traction engines and the like, subject to very definite restrictions; and the general effect of the law was to render it impractical to use, with any advantage or convenience, any self-propelled vehicle of a light kind.

Each locomotive had to be driven or conducted by three persons, one of whom had to precede the locomotive on foot at a distance of twenty yards; the maximum speed was limited to two miles an hour in towns and villages and to four miles an hour elsewhere; and there were other conditions, as to construction and use, with which a motor could hardly conform."

[16] *The Locomotives on Highways Act* 1896; Section 1: Exemption of Light Locomotives from certain statutory provisions.

[17] *Ibid* at 1(1)(b).

provisions, considered more suitable to their character[18], prohibited them from: (i) drawing more than one vehicle (a single 'trailer'); (ii) weighing more than three tonnes unladen or four tonnes with trailer; and (iii) emitting visible smoke or vapour, except due to a temporary or accidental cause[19].

While also regulating lights, a bell, new speed limits and fines of up to ten pounds, LOHA made no provision for registration of vehicles or licensing of drivers. However, LOHA resulted in significant growth in the use of vehicles in the UK, which led groups from the motor vehicle industry and general public to call for further changes to the law. These included requests to: (a) remove the three tonne weight limit to enable development of larger vehicles for commercial haulage purposes; (b) legislate an approach for identifying individual vehicles in order to ensure absent offenders could be brought to justice for driving offences; and (c) increase the regulated speed limits where no danger to others would arise from a higher speed[20]. Parliament developed *The Motor Car Act* 1903 (MCA) to meet the broader intent of these requests. However, the MCA was originally passed with a finite lifespan. All provisions would cease to have effect on the last day of 1906 absent further intervention from Parliament[21].

The MCA resolved requests for a means to identify individual vehicles through the requirement for registration of, and affixing of identification to, the newly titled *motor car*[22]. However, MCA went one step further by also requiring identification (and taxation) of the driver. This was to be achieved by application for and, on request by a police constable, production of a *driver's license*[23] - a paper document of the type shown in **Figure 2** which could be purchased from the local county or borough office for the fee of five shillings[24]. Practical skills tests were introduced in 1935, with a Mr R. Beere being the first UK citizen credited with passing the new driving test, having received the certificate shown in **Figure 3**[25].

---

[18] Selby, W., Harrel, D., Forwood, W., Henry, E., Mure, W., Monro, H. & Bigham, C. (1906). Volume 1: Report of the Royal Commission on Motor Cars. *United Kingdom House of Commons*. Cd. 3080; at 6(b).
[19] 15 at 1(1).
[20] 17 at 8(a-c).
[21] *The Motor Cars Act* 1903, Section 21. As a result of recommendations in the Report of the Royal Commission on Motor Cars, the Act was extended beyond 1906.
[22] *Ibid* at s2.
[23] 20 at s3.
[24] The Driver License fee of five shillings was fixed by s3(2) of the Act.
[25] https://www.gov.uk/government/publications/history-of-road-safety-and-the-driving-test/history-of-road-safety-the-highway-code-and-the-driving-test

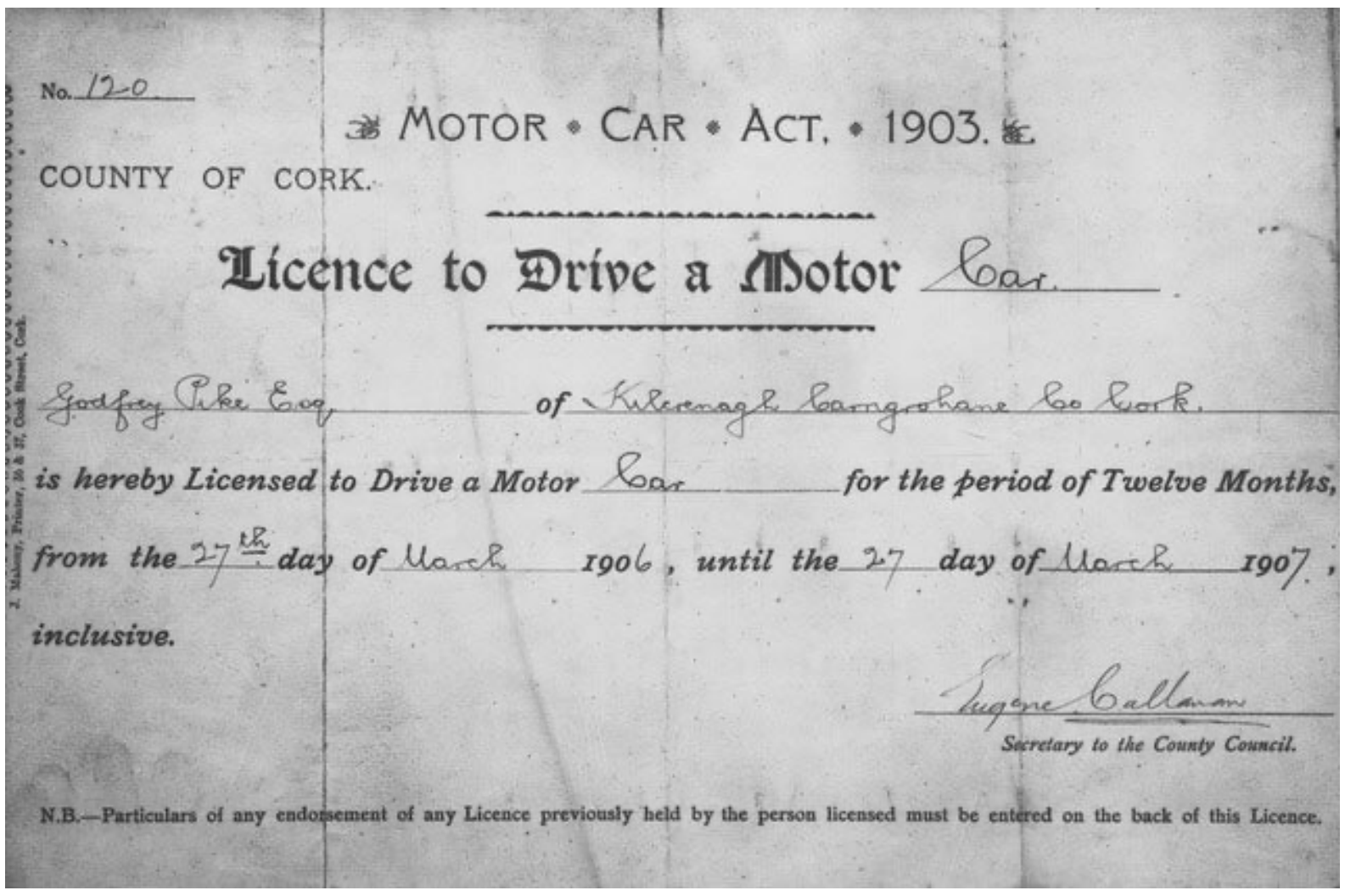

No. 120

MOTOR • CAR • ACT. • 1903.

COUNTY OF CORK.

Licence to Drive a Motor Car.

Godfrey Pike Esq, of Kilcrenagh Carrigrohane Co Cork.

is hereby Licensed to Drive a Motor Car for the period of Twelve Months,

from the 27th day of March 1906, until the 27 day of March 1907,

inclusive.

Eugene Callanan

Secretary to the County Council.

N.B.—Particulars of any endorsement of any Licence previously held by the person licensed must be entered on the back of this Licence.

***Figure 2****: Example of a 1903 Act motor car driver license  from the County of Cork, UK.*

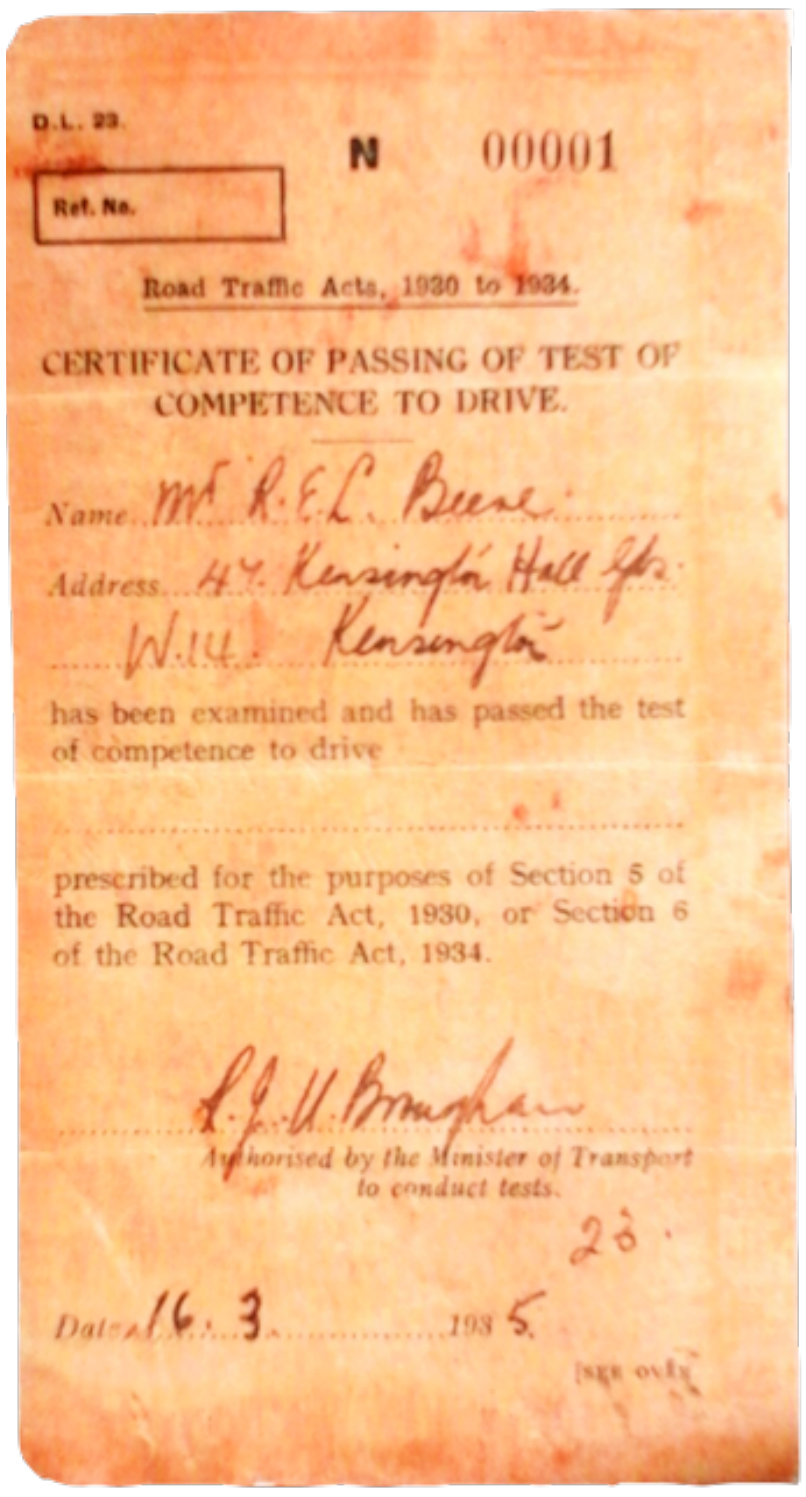

D.L. 23.

N 00001

Ref. No.

Road Traffic Acts, 1930 to 1934.

CERTIFICATE OF PASSING OF TEST OF COMPETENCE TO DRIVE.

Name Mr R.E.L. Beere

Address 47 Kensington Hall Gds. W.14. Kensington

has been examined and has passed the test of competence to drive prescribed for the purposes of Section 5 of the Road Traffic Act, 1930, or Section 6 of the Road Traffic Act, 1934.

Authorised by the Minister of Transport to conduct tests.

23.

Date 16.3 1935

***Figure 3:*** *Mr Beere's Test Certificate*[26]

## 2.1 Summary

The 1890's and early 1900's saw significant social and regulatory change. Petrol, and even some electric automobiles, went from concept to reality and progressed on to become patented, mass-produced and commonplace. Vehicle registration and affixation of what we now come to call *number plates* also arose to enable identification and prosecution of absent scofflaws. Driver licensing itself was initially little more than a method for addressing the dominant impression of the day among affluent vehicle owners that driving was a *right*, through the imposition of a tax in exchange for a permit while authorities generated a record those who may now be using the recently arrived *motor car*. Finally, the dominant purpose for registration and requirement to affix visible number plates to vehicles has largely remained the same since that time. However, as we will see in the next section, the physical medium and in many cases purpose of driver's licenses has changed considerably during the last several decades. **Table 1** summarises the introduction of vehicle registration and driver licensing, testing and training for the early-adopter countries included in this review.

[26] Sourced from: https://www.gov.uk/government/publications/history-of-road-safety-and-the-driving-test/history-of-road-safety-the-highway-code-and-the-driving-test

***Table 1:*** *Regulatory requirement for driver licensing and vehicle registration - year of introduction*

| Country | Vehicle Registration[1] | Driver's License | Driver Testing | Driver Training |
|---|---|---|---|---|
| **Germany** | 1896 | 1909 | 1903 | 1903 |
| **France** | 1893 | 1893 | 1893 | 1917 |
| **United Kingdom** | 1903 | 1903 | 1935 | |
| **United States** | 1899 | 1899 | 1908 | 1930's |
| **Australia** | 1906 | 1895 | 1911 | |
| **New Zealand** | 1905 | 1902 | 1931 | |
| **Canada** | 1903 | 1909 | | |
| **Switzerland** | | | 1890 | 1905 |

[1] Including requirement for an identifying mark or number plate

# 3. Evolution of the Driver's License

Driving licenses have continued to evolve not just physically, but also in the process of their acquisition, types and usage context. We now consider the evolution of driver's licenses from those early paper documents that simply permitted the use of motor vehicles on public highways to something that, in some countries, stands in lieu of a single federated identity card as the ubiquitous government-issued identity document.

## 3.1 Physical Evolution

We have limited this review to seven countries to provide a suitable scope for contrast and comparison. While not all countries regulate driving in exactly the same way, there are sufficient data points for these early-adopter countries in **Figure 4** to observe general consistency as driving licenses transitioned through different physical (and eventually fully digital) media.

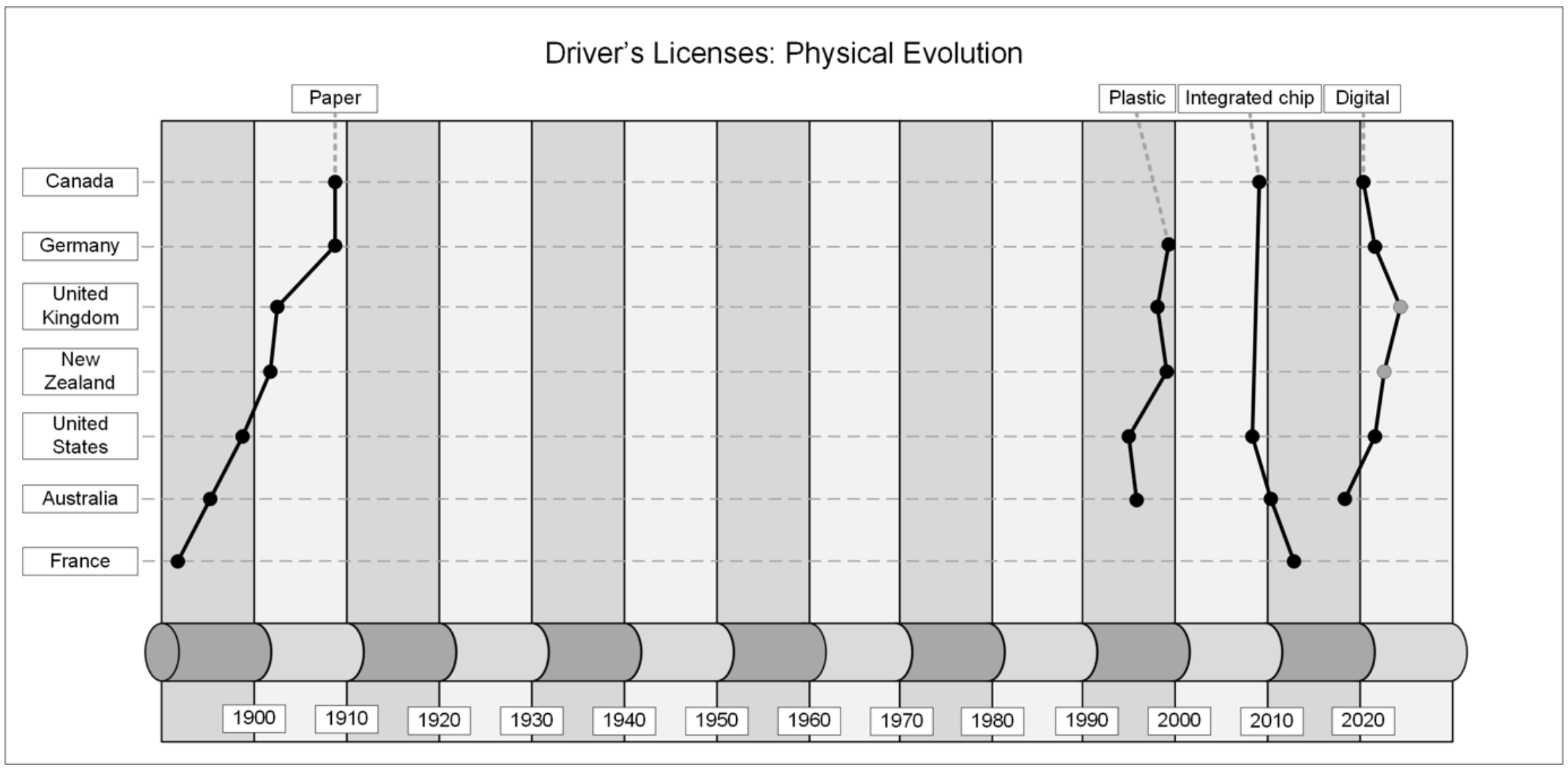

***Figure 4:*** *Comparative physical evolution of Driving Licenses*

Every country we reviewed (and we would suggest the majority of countries that were early motor vehicle adopters) began with a paper-based driving permit - this includes those countries that printed permits on cardstock. Paper-based licenses persisted in most countries until the 1990's. While paper licenses in countries like Australia (see **Figure 5**) contained very little information, Chauffeur's licenses in New York state from as early as 1911 contained a monochrome photograph and physical description of the license holder (**Figure 6**). Full colour passport-style photographs were affixed for many years to French paper driver's licenses (**Figure 7**).

The move to plastic credit-card style driving licenses, often with clear coatings containing holographic or tessellated images and other anti-counterfeiting technology, was promoted as a step towards preventing identity theft and document fraud (Kemp, Towell & Pike, 1997, Dobrowski et al, 1973). The United States of America, closely followed by Australia during the year after, were the first to roll out plastic laminated driving licenses. Similar fraud prevention reasons were also given for the more recent implementation of biometric chips in countries like the United States and Canada, whereby images and other data about the license holder are stored in an integrated smart chip (Adeoye, 2010). For example, these *enhanced driver's licenses* (EDL) enabled Canadian and United States citizens to pass through several border crossings between the two countries without the need to carry a full passport. However, social distancing anxieties during 2021 that halted in-person interviews necessary to acquire the EDL coupled with the fact that uptake had never been sufficient to justify the added public expense, led the Canadian Government to recently shutter the EDL program[27]. During the late 2010's several countries proposed, trialled, and in a few cases deployed[28] digital driving licenses that the holder can download using a government-provided 'app' and store locally in their smartphone device. Countries including Australia, Canada, Germany and the United Kingdom have described a future, beyond 2024, where the Government want the only form of driving license being issued to be digital.

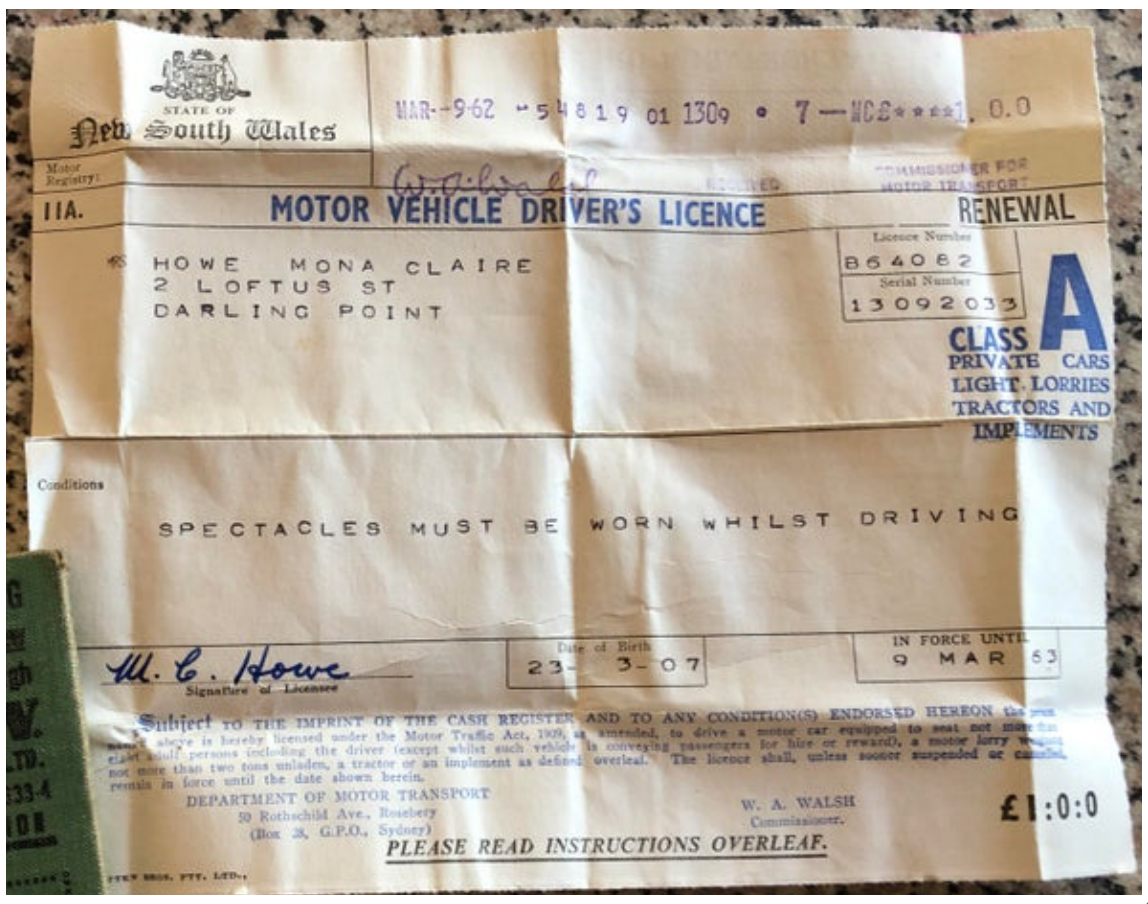

***Figure 5:*** *NSW Paper Driver's License issued 9 March 1963*[29]

[27] Zussman, R. (2021). British Columbia is getting rid of enhanced driver's license and enhanced identification card. Last accessed 31 January 2022. Sourced from: https://globalnews.ca/news/7585720/phased-out-bc-enhanced-licence/

[28] Colorado was one of the first in the United States to trial a digital driver's license in 2017 - then a QR-code based DigiDL App and since 2021 in full deployment as the *Colorado Digital ID* - a state-issued identification and digital driver's license synchronised to the holders smartphone after registration of the myColorado App.

[29] https://i.redd.it/h4jojrbdwmt31.jpg

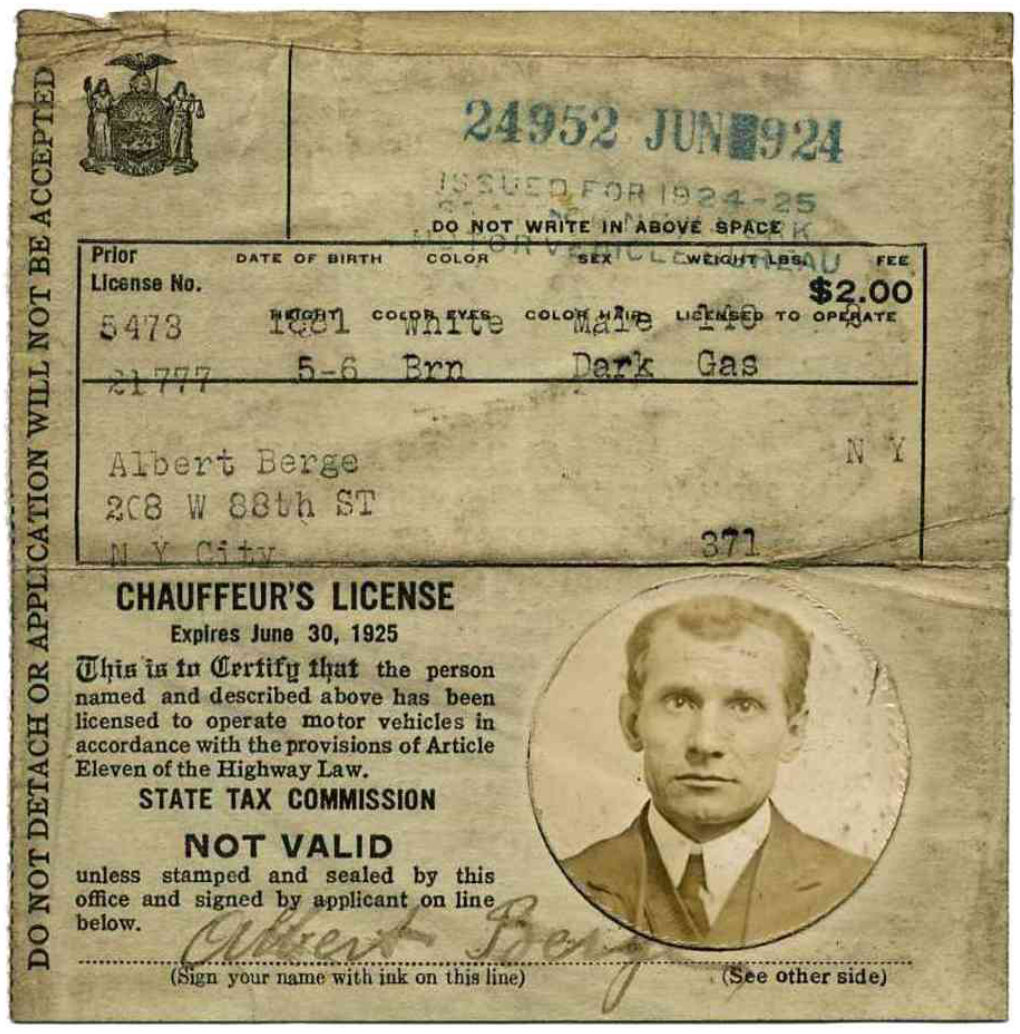

*Figure 6:* *New York Chauffeur's License circa 1924*[30]

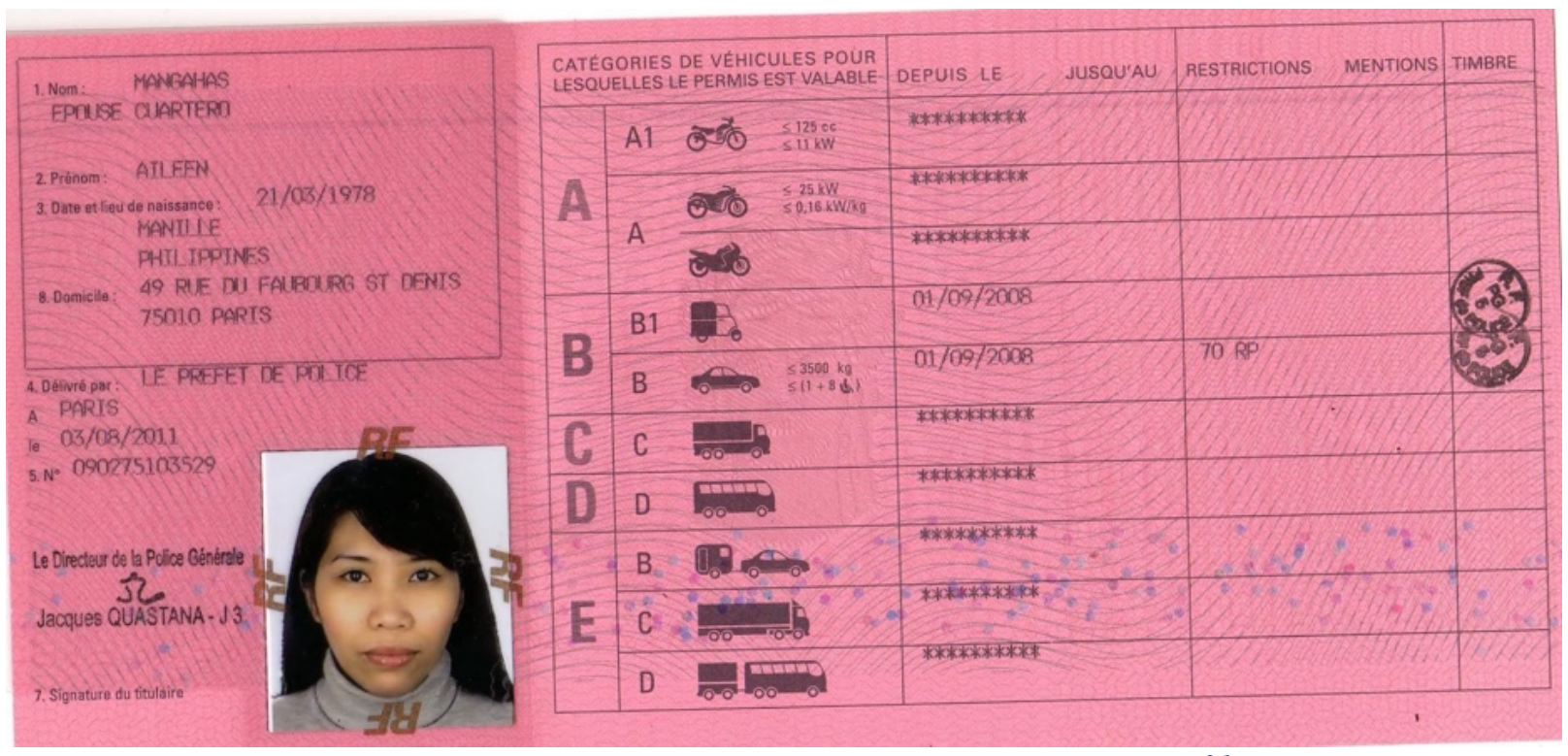

*Figure 7:* *French Driver's License circa 2011*[31]

## 3.2 Contextual Evolution

The contextual basis and use case for driver's licenses has changed considerably during the last century. Like Benz's driving permit most early 'licenses', including those of New Zealand and the United Kingdom, simply identified the driver/owner who paid the fee to permit the vehicle's use on public roadways (Finnerty, 2015; Pawson, 2014). While the French moved fairly quickly to change the context of driver's licenses from a *permit to use* to a *certificate of competency*, for countries like the United Kingdom it took another 40 years for the proficiency requirement to become law. Further, even though more than 100 countries have enacted some form of compulsory *State-issued Identity Card*[32] (SIC) the driver's license has consistently become the *de facto* SIC, especially where compulsory SIC have failed to gain popular acceptance[33]. For example: Australia, the United Kingdom and many North American states, including Colorado[34].

---

[30] http://graphics8.nytimes.com/packages/images/multimedia/bundles/projects/2013/Licenses/1924op_2x.jpg

[31] http://2.bp.blogspot.com/-wLXu-0r4haw/UtgJ8W10DMI/AAAAAAAAFbg/AsLG1NaY2Ro/s1600/licence.jpg

[32] https://web.archive.org/web/20110903074029/https://www.privacyinternational.org/article/id-card-frequently-asked-questions

[33] While there has been vigorous public outcry in countries like Australia and the UK regarding introduction of SIC on multiple occasions, it has generally been focused on issues of physical acquisition, cost, daily use or control of the individual and has ignored the larger issue of the central or 'joined-up' database that usually comes with such identity systems that actually poses the real threat. See: https://privacy.org.au/about/history/davies0402/

[34] The *Colorado Digital ID* in the State-run myColorado App acts as both a digital state-issued ID and digital driver's license.

## 3.3 Test and Type Evolution

During the more than a century of motor vehicles, there have been significant changes to the types and degree of proficiency testing required to receive a driver's license. This culminated in many jurisdictions with introduction of phased or *graduated* driver's licenses (GDL). GDL are intended to expose the new driver to the increased responsibilities such as driving at night or highway speeds - only after they have attained sufficient experience in less demanding conditions. Further, many different types of driver's licenses are available: from those that allow only the use of 50cc motor scooters through to specialist licenses for drivers of large oversized load-carrying articulated semi-trailer trucks. This section discusses these changes, while **Appendix A** describes the separate key historical events identified in each country with links to source materials.

*Practical driving tests* were introduced in the reviewed countries over a period of five decades, and often in response to increasing numbers of traffic accidents and road casualties[35]. As traffic law and road rules became more extensive and detailed, the concept of *Theory Testing* involving verbal[36], written or multiple choice questions, or lottery-style scratch-off exam cards[37] evolved. The United Kingdom was one of the later adopters of written theory tests, which were very quickly adapted into the current computer testing regime. With some proponents claiming test scores are indicative of driver skill and crash risk, the *Hazard Perception* (HP) test first proposed more than 50 years ago [38] has only recently become part of the mandatory driver testing process in some countries. Current approaches to HP testing use situational and simulated scenarios to evaluate the driver's *situational awareness* and corresponding responses to potential hazards (Underwood, Crundall & Chapman, 2011). **Figure 8** provides a visual reference of the temporal variation in evolution of proficiency requirements in motor vehicle early-adopter countries from the first practical driving tests through to the recently introduced hazard perception (HP) testing.

---

35 https://www.2pass.co.uk/history-driving-test.htm and https://drive4lifeacademy.co.uk/history-driving-test-infographic/

36 Prior to the formal introduction of written theory tests, it was common for an examiner to ask the candidate a number of verbal theory questions to ascertain knowledge of the road rules before commencing the practical driving test.

37 Until computer testing was adopted across New Zealand, new drivers and those converting an overseas license, completed lottery-style scratch card tests as their initial theory exam. Each test card included a variety of road rule knowledge (scratch off the square corresponding to the speed limit in a school zone between 2:30 and 4:00pm) and visual situation-style questions (an image might display three cars identified with alphabetical letters at an intersection and ask the test taker to scratch off the letter corresponding to which car had right of way to proceed first).

38 Horswill, M. (2016). Hazard Perception Tests, in Fisher, D., Caird, J., Horrey, W., & Trick, L. (Eds.). *Handbook of teen and novice drivers*. CRC Press: Boca Ratan.

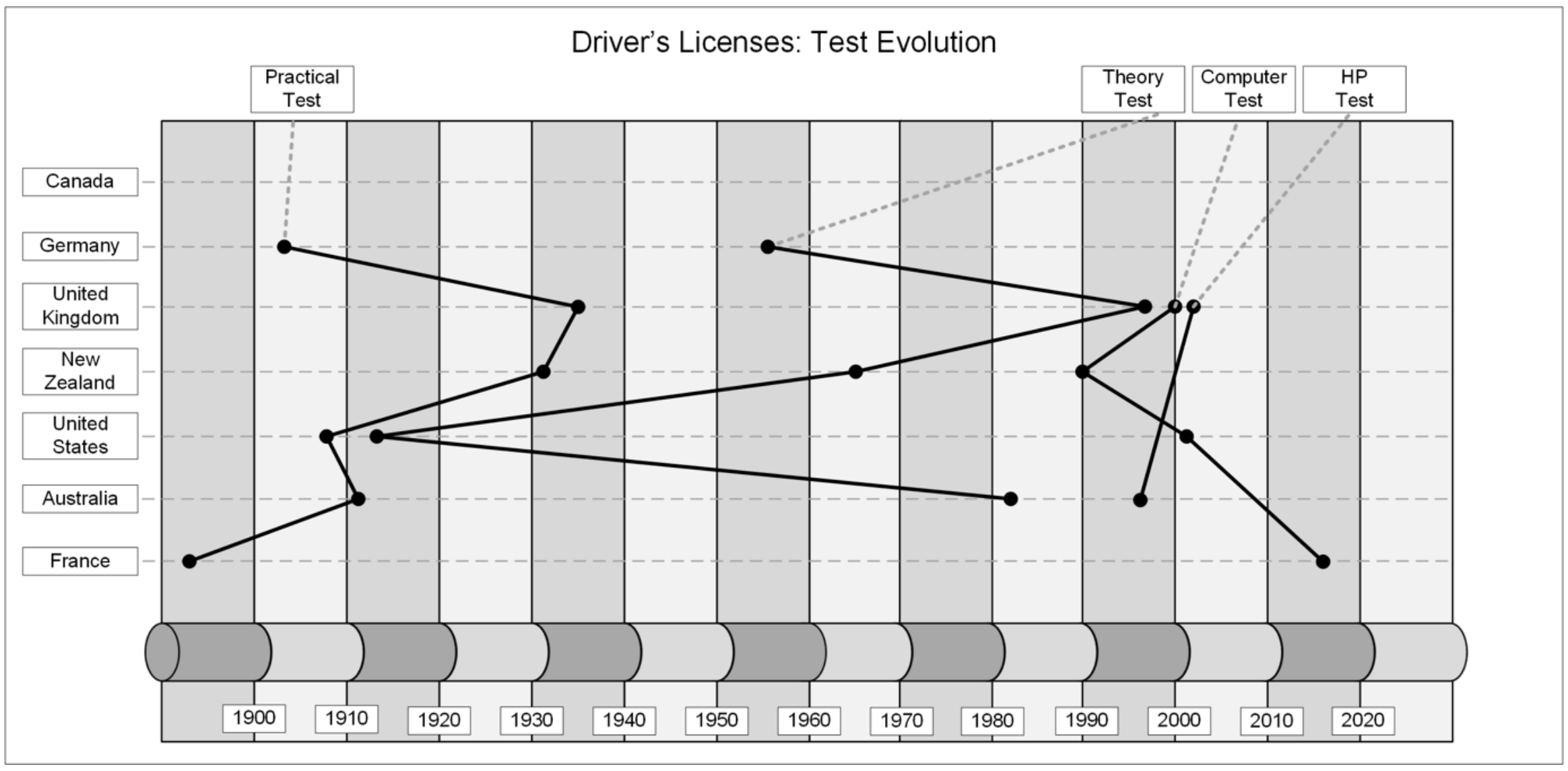

***Figure 8:*** *Driver's Licenses - Test evolution*

We have already seen that the earliest unrestricted (or full) driver's licenses were purchased for a fee rather than earned through acquisition and assessment of skill. As motor vehicles became increasingly more popular, new approaches were sought to slow the concomitant rise in accidents and casualties. Provisional (learner) and probationary (restricted) licenses were introduced that were eventually encapsulated as phases of the *Graduated Driver's License* (GDL) (Williams, McCartt & Sims, 2016). Today in most countries the new driver will undergo some period of supervised instruction as a learner (L) driver followed by testing to demonstrate mastery of the necessary skills and knowledge. The novice driver then undertakes a period of 1-3 years on a restricted or probationary (P) license which, on successful completion and in some countries only after more testing, will see them graduate to holding an unrestricted or full driver's license. **Figure 9** indicates when the same early-adopter countries introduced learners, probationary and graduated driver's licenses.

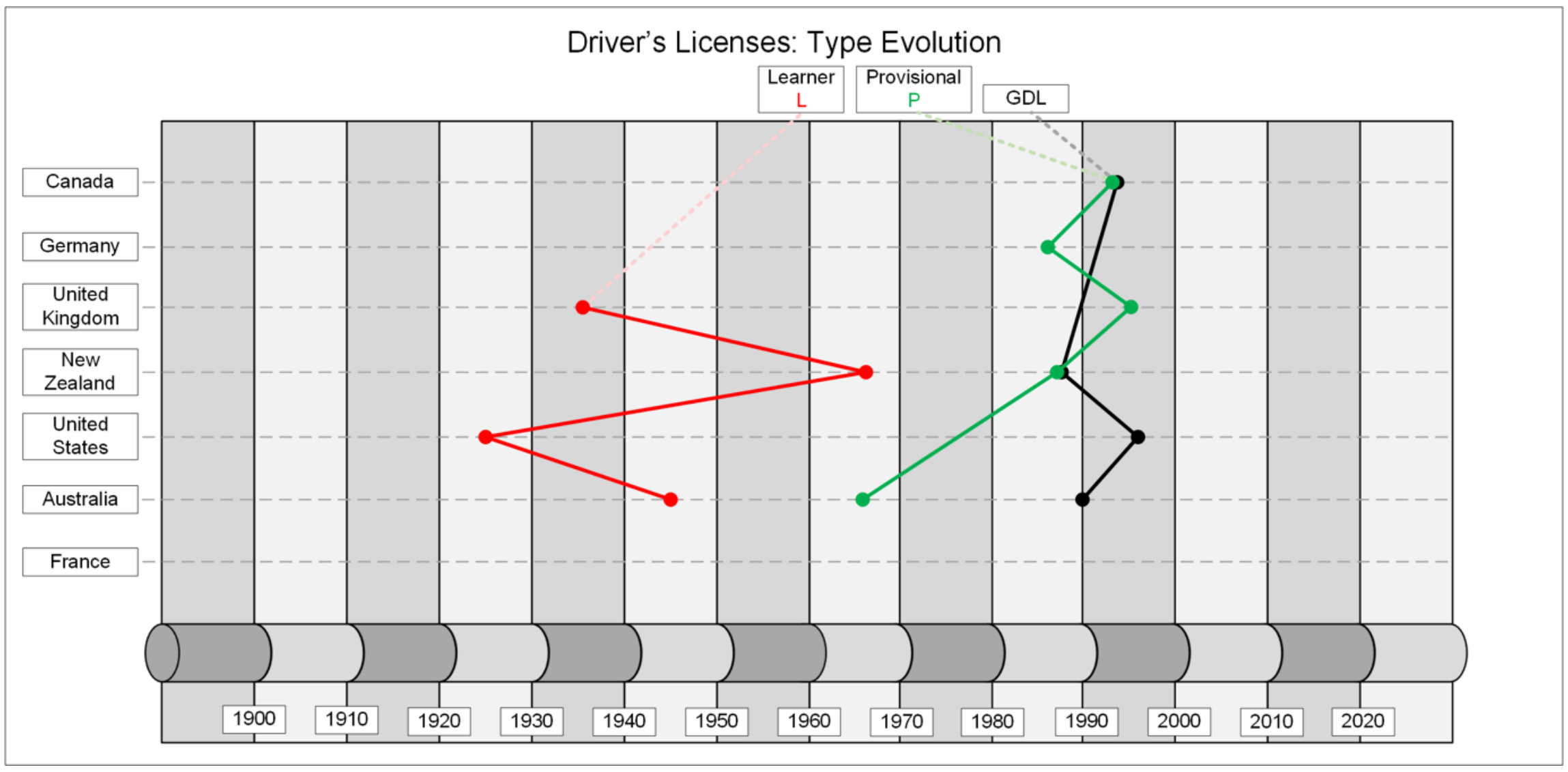

***Figure 9:*** *Drivers Licenses - Type evolution*

## 3.4 Summary

Just as automobiles have changed over the last century, so too has the driver's license. New drivers in the early 1900's simply paid a fee and received a piece of paper, while today's new drivers undergo multiple theory and practical proficiency tests to initially receive a limited right to drive under constrained rules on what is termed a *restricted* or *probationary* license. Changes in vehicle size, type and speed have also led to different classes of vehicle - so while those early licenses permitted use of motor vehicles and traction engines, today's license holders require a different class of license to drive different types of vehicles - motorcycles, passenger cars, articulated goods vehicles and so on. In some countries new drivers are now required to undergo testing using Hollywood-style animated video sequences known as *hazard perception* (HP) tests - very reminiscent of how airline pilots are trained and type-rated on commercial aircraft. Another element that is similar to the process for licensing pilots is the need in countries like the UK and Australia for novice drivers to complete a prescribed number of hours driving experience prior to undertaking the test to receive their license. The key questions that remain to be discussed in the next section are: *what will happen to the driver's license as digital takes over* and *as automobiles edge ever-closer to true autonomy?*

# 4. The Future of Driver's Licenses and Vehicle Registration

## 4.1 The Digital Driver's License

As seen in **Figure 4** almost every country in this review has already implemented, or has set a date to introduce, the digital driver's license (DDL). Also known as the *mobile driver's license* (mDL), the DDL concept has become one of a number of *digital citizen* projects currently being deployed by governments all over the world.

Leaving aside less tangible situations[39] that could arise when verification systems or network access are inaccessible[40], the potential for privacy and security issues must not be overlooked. Companies that develop digital identification and driver's license systems for State authorities play on public fears by overstating privacy risks of your existing physical driver's license in common situations such as using it as proof-of-age ID at a night club, while endorsing the software engine they sell to governments by understating any potential privacy and security risks that digital apps on smartphones are commonly seen to have, and claiming their product has impregnable security and unassailable privacy[41]. This style of privacy and security claim was also made about the software engine (in that case described as an *advanced programming interface* or API) supplied by Apple and Google for the recent contact tracing smartphone Apps. In each case a State authority produced a locally customised contact tracing App for citizens to download, but numerous privacy breaches occurred involving both the State-customised App and the underlying Google/Apple API, including: (a) log files containing personal information that users were told in press releases would not be stored locally but was actually found to be accessible in plain text by other Apps on the mobile device[42]; (b) the ability for State intelligence and law enforcement to identify and freely collect what was supposed to be securely encrypted App data while it was in-transit on the internet[43]; and, (c) State authorities profiting by selling App user's location data using the device's Google Advertising ID[44]. These breaches of public privacy and trust underline why every citizen should be wary of not just the push to move our identification and driver licensing into these digital platforms, but also, as several governments have proposed, the merging of digital identification and other services like vehicle registration, contact tracing, vaccine certification, medical records and even access to money[45] into a single or government-controlled smartphone App[46]. It is ironic that while condemning the pervasiveness and perceived excessive use of smartphones[47], and in certain controlled situations like driving[48], any use at all - with the introduction of these State-controlled App infrastructures and digital IDs the same governments also make it almost impossible to interact with their services and engage in those controlled activities without one.

[39] Consider the situation for drivers in jurisdictions where failing to present on request a valid license while driving is an offence, where the only driver's license the individual is carrying is the digital one and verification systems are unavailable at the time. Unsympathetic police who are 'only following the letter of regulation' may issue on-the-spot fines or potentially detain such drivers until their identity and right to drive can be verified.

[40] On their first day of production deployment, both the New South Wales (Australia) and Norwegian digital driver's license platforms crashed as a result of significant under-estimation of system specification and network load. https://www.newsinenglish.no/2019/10/01/digital-drivers-license-crashes/ and https://www.dailymail.co.uk/news/article-7624545/NSW-Digital-driving-license-app-crashes-thousands-try-log-on.html

[41] Thales Gemalto provide a document on their website, linked to from the American Department of Motor Vehicles (DMV) and National Institutes of Science and Technology (NIST) that purports to be 'the facts' about digital driver's licenses, but which may only be potentially misleading marketing spin for promotion of State-produced apps using Gemalto's technology: https://www.thalesgroup.com/sites/default/files/gemalto/Gemalto_DDL_FAQ_Sheet.pdf

[42] https://themarkup.org/privacy/2021/04/27/google-promised-its-contact-tracing-app-was-completely-private-but-it-wasnt

> Paula Mathews, an attractive petite 20-year-old university student, is driving home late one evening when she is pulled over by a police officer. The officer requests her driver's license, which is only available from a government-provided App on her smartphone. Paula unlocks her smartphone and opens the app, which displays an image similar to the one shown in **Figure 10**. The officer makes a show of trying to scan the QR code, complains that it is not working and tells Paula to wait in her car while he takes her device in order to verify the license on his squad car's computer.

***Text Box 1:*** *Digital Driving License Privacy Scenario*

Contemplate the digital driver's license privacy scenario proposed in **Text Box 1**. It is said that people carry their entire lives in their smartphones: that *we have become human snails carrying our home in our pockets*[49]. Most people have a multitude of private information including their location history, text messages and emails that they would not wish read by anyone else. Almost half of all smartphone users[50], some as young as 14[51], have sexually explicit *selfie* images on their devices. Few, if any, would want to hand over their unlocked smartphone to a stranger (Egelman et al, 2014), especially a police officer (Flynn, 2021).

---

43 https://techcrunch.com/2020/11/24/australia-spy-agencies-covid-19-app-data/ and https://7news.com.au/news/western-australia-police/wa-police-accessed-contact-tracing-data-c-3118713

44 https://www.fastcompany.com/90508044/north-dakotas-covid-19-app-has-been-sending-data-to-foursquare-and-google - note that the Care19 app from North Dakota is still available on both the Android and Apple App stores, and on each identifies that it uses the Exposure Notification API (aka the Apple/Google API) - https://apps.apple.com/us/app/care19-alert/id1513945072

45 The Bank of England have proposed replacement of the Great British Pound (GBP) with a government-controlled digital currency that "*could be programmed to ensure it is only spent on essentials, or goods which an employer or Government deems to be sensible*". https://www.telegraph.co.uk/business/2021/06/21/bank-england-tells-ministers-intervene-digital-currency-programming/

46 For example: the Government of South Australia have already merged vehicle registration, driver's licenses (including demerit points and fines), some medical records and medical self-assessment, firearms licenses, real estate registrations, covid passports and contact tracing, and other digital passes into their mySA App (https://my.sa.gov.au/ and https://www.covid-19.sa.gov.au/travel/interstate-travel). In order to process Covid Passports, the United Kingdom's NHS App was found to have been extended for collection and storage of data on vehicle registrations, criminal convictions and a range of other personal information unrelated to whether or not the subject individual had received covid injections (https://www.express.co.uk/life-style/science-technology/1437931/NHS-app-personal-data-Covid-vaccine-passports).

47 In 2018 France passed a law that banned possession and use of smartphones on school grounds. Australia's New South Wales government instituted similar policies that also incorporated other smart devices like smart watches in December of the same year. In June 2021 Nigeria blocked smartphone and computer access to Twitter claiming the social media platform was being used to propagate anti-government sentiment and coordinate mobile anti-government protests. Multiple governments have also either passed anti-encryption technology back-door laws, or are funding research into encryption workarounds for use as a 'lawful government investigation technique'.

48 Most governments have enacted laws prohibiting the use of smartphones while driving, with significant fines and demerit points as penalties for those identified using one - whether to make a phone call or compose a text message, or to take photographs of a traffic incident or change music.

49 Recent studies on smartphone use have suggested that the smartphone is possibly the first object to challenge the house and workplace in terms of the time we dwell in it during our waking hours. Miller, D., Rabho, L. Awondo, P., de Vries, M., Duque, M., Garvey, P., Haapio-Kirk, L., Hawkings, C., Otaegui, A., Walton, S., & Wang, X. (2021).

Governments and law enforcement agencies have actively sought to compel the unlocking of private citizen's digital devices for at least the last decade. In 2018 Australia and New Zealand enacted legislation allowing them to not only search digital devices at the border, but also to compel disclosure of passwords with large fines for non-compliance[52]. Recent law changes in some Australian states made it a criminal offence to refuse to supply passcodes where police have compelled production[53]. Often, very little justification is required to instantiate a search of your devices[54], and there have been several very public cases where people have refused to provide passcodes[55]. However, when people have complied and provided passcode, fingerprint or face-ID to unlock the device, not all officers have acted with decency or integrity. For example, some officers have sent and replied to messages using the person's device and then deleted evidence and sought to deny their malfeasance[56], while others have forwarded nude selfies found on attractive female suspect's phones to themselves or other officers[57].

---

The global Smartphone: Beyond a Youth Technology. UCL Press. https://discovery.ucl.ac.uk/id/eprint/10126930/1/The-Global-Smartphone.pdf

[50] Lee, M., Crofts, T, McGovern, A., & Milovejovic, S. (2015). Sexting and young people: Report to the Criminology Research Advisory Council. Grant: CRG 53/11-12. Last accessed 2 February 2022. Sourced from: https://www.aic.gov.au/sites/default/files/2020-05/53-1112-FinalReport.pdf

[51] Donald S. Strassberg, Ryan K. McKinnon, Michael A. Sustaíta, Jordan Rullo. Sexting by High School Students: An Exploratory and Descriptive Study. *Archives of Sexual Behaviour*, 2012; DOI: 10.1007/s10508-012-9969-8

[52] The *Customs and Excise Act* 2022 (NZ), Section 228 allows customs officers to search any device where they have reasonable belief that the device has been, or may be, involved in a criminal offence. This includes the power to compel passwords or other assistance necessary to facilitate access to the device, to take a copy of the device, or to retain the device. Australian border officers have a similar power under the *Customs Act* 1901.

[53] Where Queensland police have received a search warrant under Section 154 of the *Police Powers and Responsibilities Act* 2000 (PPRA) covering a digital device, they can compel production of the passcode to unlock or decrypt the device. Refusal carries a potential 5year prison sentence. However, as seen in *Barbaro v Queensland Police Service* [2000] QDC 39 - if founded and legitimate, legal privilege is at least one 'reasonable excuse' falling within the exception of Section 205A of the *Criminal Code 1899* (QLD). In another recent judgement (*Wassmuth v Commissioner of Police* [2018] QCA 290) in obiter, North J recognised that it may be possible that exercising ones right to silence and not acknowledging ownership or knowledge of how to use the device could also be sufficient to defend a prosecution for failure to comply.

[54] https://bccla.org/our_work/electronic-devices-privacy-handbook-a-guide-to-your-rights-at-the-border/

[55] https://www.nbcnews.com/news/us-news/give-your-password-or-go-jail-police-push-legal-boundaries-n1014266 and https://arstechnica.com/tech-policy/2020/02/man-who-refused-to-decrypt-hard-drives-is-free-after-four-years-in-jail/ and https://www.cnet.com/news/man-charged-for-refusing-to-give-up-phone-passcode-to-canadian-border-agents/ and https://www.washingtonpost.com/technology/2019/04/03/apple-employee-detained-by-us-customs-agents-after-declining-unlock-phone-laptop/

[56] A 22 year old man sued the Australian Government after a Border Force Officer detained him, demanded passcodes to his personal devices and, after taking the devices into another room ostensibly to search them, used the device to send and reply to text messages - after which he attempted to delete the messages from the phone and only admitted his unlawful actions after investigators from the Integrity and Professional Standards Department began investigating https://www.smh.com.au/national/nsw/legal-action-after-border-force-officer-secretly-texted-on-passengers-phone-20160219-gmy8c3.html

[57] A police officer in California was prosecuted after he gained access to at least two female suspect's phones and forwarded nude selfies from their phones to himself or other officers. https://www.cnet.com/news/cop-charged-with-stealing-nude-photos-from-suspects-iphone/. A Minnesota officer was sacked after texting himself nude selfies from the phone of a young woman involved in a traffic accident. https://www.nbcnews.com/news/us-news/former-minnesota-trooper-pleads-guilty-texting-himself-nude-photos-woman-n1270341. An Australian police officer lost his job and was fined after he shared nude selfies he found during the search of a woman's phone on Facebook - identifying her and saying that the images came from her phone. https://www.theguardian.com/australia-news/2019/feb/19/police-officer-who-stole-intimate-images-from-womans-phone-avoids-jail

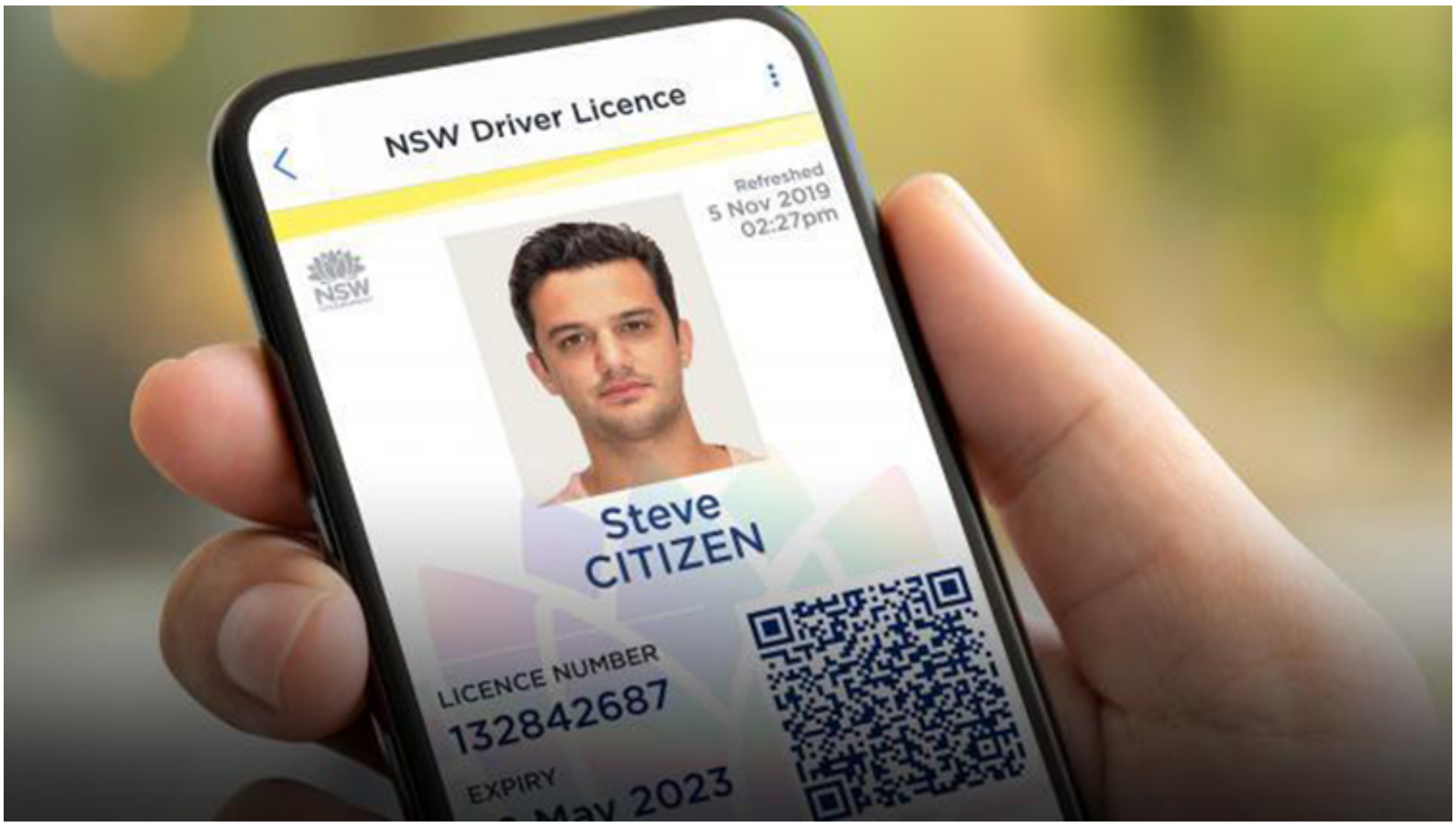

***Figure 10:*** *A digital driver's license*

Consider: The police officer is walking back to his patrol car with Paula's unlocked smartphone ostensibly to validate her DDL. Would your willingness to adopt digital driver's licenses change if Paula was your daughter? Or if you knew this police officer had previously been censured for looking up the personal details of people they were attracted to in police databases?[58] Or had previously harassed or stalked young women?[59] Or if you knew that Paula was a social work student currently on placement in the women's refuge where the officer's wife had recently fled with her children?[60] Or that Paula was incorrectly identified as a person of interest in a crime and the officer was using the ruse of a 'pretext' stop to see if there was anything in her phone that might support a later search or arrest warrant?[61]

---

[58] Many police officers have been reprimanded, censured, sacked and even prosecuted for unlawfully accessing personal details of people they were attracted to or dating using police databases accessible via police computer systems. See: https://www.dailytelegraph.com.au/newslocal/fairfield-advance/matthew-price-fairfield-cop-slept-with-woman-after-looking-her-up/news-story/ccb087293ce81924cfa6d2f40da6c3dd and https://www.dailymail.co.uk/news/article-9748915/Female-police-officer-sacked-using-force-computer-look-details-prospective-date.html and https://www.abc.net.au/news/2019-02-01/officer-used-police-computer-to-look-up-tinder-dates/10771958

[59] Lanark Special Constable Ryan Reid was jailed for 15 months in 2015 after evidence showed he had stalked a string of young women after accessing their personal details in confidential police computer systems and in some cases, sharing them with friends. https://www.dailymail.co.uk/news/article-2931746/Police-officer-stalked-string-young-women-jailed-15-months.html

[60] Over a three-year period 31 police officers in Australia's New South Wales were prosecuted for domestic violence offences. In one case, when reporting the alleged abuse by her uniformed partner, the female victim was threatened with retributory action and potential prosecution if she continued to report the violence of her '*work-stressed'* abuser. https://www.abc.net.au/news/2021-05-10/nsw-police-officers-charged-with-domestic-violence-2020-victims/100114114

[61] Known in America and Australia as 'pretext' traffic stops, such stops have been used to target people of interest or protected groups within the community. The officer pulls the driver over on a pretext - a blown light bulb, for driving too cautiously, or by using a specious claim that they 'reasonably suspect that the vehicle contains a weapon or drugs' - which is a legislative loophole allowing warrantless searches in some jurisdictions - upon which the officer uses the stop to: (a) visually search the inside of the vehicle; (b) demand that the driver open the rear storage compartment of the car; or (c) to identify other occupants in the vehicle.

The shift to digital identification and driver's licenses might seem *in vogue* given our current generation of young people's proclivity for adopting *smart* technology. The Australian[62] and UK[63] governments are both developing legislation under ostensibly similar draft digital identity Bills. Such digital identity laws will usher in an even more invasive digital surveillance age where the government is more able to centrally monitor every online interaction. Having already begun the process with introduction of *Covid passports*, track and trace *digital check-in*, and *DDL*; these digital identities will be central to every single person's life and have been described publicly as *a license to live*[64]. However, the risks involved in shifting to digital-only and any potential ulterior motives of some government and law enforcement actors should give us pause. Do we want to be completely reliant on what has and will always be poorly designed, inherently unreliable and easily exploited State-run technology?[65] What effect would it have on your ability to access other services that require identification if your DDL is your primary State-issued identification and the App service is either not working, buggy or your access has been restricted? How long will it be before, as was suggested by Bank of England officials for the United Kingdom's recently promoted digital currency[66], the government starts making decisions about when and where you can access or use your DDL, and hence drive - restricting it for disciplinary, social, supposed environmental or other reasons?

## 4.2 Driver's Licenses in the Autonomous Future

Writers and researchers have variously suggested that fewer[67] or greater[68] numbers will drive in future. They also predict we either won't need a driver's license[69], or that the nature of driver's licenses will change from certifying driving proficiency to certifying an ability to

---

[62] In Australia legislation has finally been made public under the deceptively titled: *Trusted Digital Identity* Bill. This 149 page draft bill requires identification of every individual when they create an account online via a centralised Government-run digital identity stack linked to all other State and Federal services, and as more organisations like telecommunications providers and retailers are brought within this legislation, it means when going online most people will cease to have the option to be anonymous and the system will allow the Australian government to collect in near real-time vast amounts of data including what is said, done or purchased almost everywhere that is required to use this digital identity protocol. https://www.digitalidentity.gov.au/sites/default/files/2021-09/Trusted%20Digital%20Identity%20Bill%202021%20exposure%20draft.pdf

[63] https://www.gov.uk/government/publications/uk-digital-identity-attributes-trust-framework-updated-version/uk-digital-identity-and-attributes-trust-framework-alpha-version-2 and https://brc.org.uk/news/the-retailer/digital-identity-what-will-it-mean-for-retailers-and-their-customers/

[64] Malcolm Roberts, Senator for Queensland.

[65] Peeters, R., & Widlak, A. (2018). The digital cage: Administrative exclusion through information architecture–The case of the Dutch civil registry's master data management system. *Government Information Quarterly*, *35*(2), 175-183.

> Describes where digital information exchange between various government actors over 16 years created a real-life situation where a woman who'd reported her car stolen was never told by police that it had been found - but the tax department, traffic regulator and others were told digitally through government data exchange services. They charged her years of registration fees, fined her for failing to perform annual roadworthiness checks, and taxed her tens of thousands of dollars on the basis that she was still the owner of the car and their erroneous belief that the car had been returned by police into her possession.

[66] See note 44.

[67] https://www.wired.com/2012/09/ieee-autonomous-2040/

[68] https://www.fastcompany.com/3054553/in-a-driverless-car-future-more-people-will-drive-because-they-dont-need-a-license

[69] See note 66 - FastCompany and https://www.wired.com/2012/09/ieee-autonomous-2040/ and: Wing, C. (2016). Better keep your hands on the wheel in that autonomous car: Examining society's need to navigate the cybersecurity roadblocks for intelligent vehicles. *Hofstra L. Rev.*, *45*, 707 at p.745.

command an AI-based self-driving vehicle[70,71]. Some have promoted introduction of *vehicle licenses*[72]: a driver's license issued to the vehicle instead of the human occupant in what would otherwise be the driver's seat. This approach could certify the complete built-up ecosystem of the autonomous vehicle type in a similar way as UNECE's WP.29[73] is intended to evaluate, verify and certify the manufacturer's approach and vehicle type's cyber security[74]. One complication would be the way licensing the vehicle as driver would cloud the legal distinction that exists between manufacturer and driver; potentially vesting legal responsibility for not just the design, materials, and construction of the vehicle, but also for *at fault* accidents that occur when the autonomous system is in control with the manufacturer. This approach to resolving liability was proposed by a quarter of autonomous vehicle researchers in one recent study[75].

Should we wonder how manufacturers might deal with situations where autonomous systems act inappropriately and serious harm or death results, we can turn to examples in other industries that have used automation in safety critical control systems. Drawing again from the aviation industry we could consider how Boeing responded to MCAS[76] issues in their 737 Max 8 and 9 airplanes. Boeing first knowingly and deliberately concealed the existence of MCAS[77], and removed all mention of the MCAS system from flight manuals and training materials[78], thus never alerting pilots that MCAS could take control of their plane in a way that was completely different to the autonomous control systems and flight characteristics of any previous 737 plane. After two airplanes crashed in similar fashion[79], Boeing denied that MCAS could be at issue, preferring instead to blame the Lion Air and Ethiopian Airlines pilots and suggesting that had they been American pilots the accidents would not have happened[80]. It took almost a full year and, only after most countries had grounded more than one hundred 737 Max

---

[70] Dr Lance Elliot seems to suggest that more of us will in future get a driver's license, but that the nature of that driver's license will change. It will be less about your driving proficiency and more about certifying you as an *AI driverless autonomous car driver* - much like how the UK driving tests at the moment require demonstration of ability to use and carefully consider the directions of in-vehicle turn-by-turn driving systems. https://www.google.com.au/search?q=more+people+will+have+drivers+licenses+with+autonomous+cars&ei=rO4AYvnFMtCagQbkprIY&start=10&sa=N&ved=2ahUKEwj5y5LPq-31AhVQTcAKHWSTDAMQ8NMDegQIARBP&biw=1722&bih=890&dpr=2

[71] Wing, C. (2016). Better keep your hands on the wheel in that autonomous car: Examining society's need to navigate the cybersecurity roadblocks for intelligent vehicles. *Hofstra L. Rev.*, *45*, 707.

Wing argues in section 4 of his work that current driver's licenses should be replaced with a specialised license to operate an autonomous vehicle.

[72] See Wing at note 71, p744.

[73] https://unece.org/fileadmin/DAM/trans/doc/2020/wp29grva/ECE-TRANS-WP29-2020-079-Revised.pdf

[74] McLachlan, S. & Schafer, B. (2021). Too many moving parts: Autonomous vehicles and the modelling of complex technico-legal systems. *Presentation at the 2021 BILETA Conference* (Online). Presentation Video: https://www.youtube.com/watch?v=sDyR66J3EJs&t=2s

[75] McLachlan, S., Kyrimi, E., Dube, K., Fenton, N., & Schafer, B. (2022). The Self-Driving Car: Crossroads at the bleeding edge of Artificial Intelligence and Law. *ArXiv Preprint.*

[76] The Manoeuvring Characteristics Augmentation System (MCAS) is an autonomous flight stabilisation computer program implemented by Boeing in 737 Max8 and Max 9 airplanes to counter for design issues and oversized engines that tended to push the airplane's nose up at high thrust.

[77] Investigations identified internal emails and documents showing Boeing staff intended to present MCAS as an enhancement to the existing Speed Trim system that assists pilots to keep the airplane in *trim* (level flight) rather than as what it was - a completely new autonomous flight handling system that required retraining for pilots and revision of flight manuals and procedures. Boeing even spoke in internal emails of "jedi-mind tricking regulators" to ensure that they could sell the plane on the basis that no new changes had been made to the airplane and thus no additional pilot training in simulators was required. https://www.marketwatch.com/story/senators-hammer-boeing-ceo-over-pilots-talk-of-jedi-mind-tricking-regulators-2019-10-29

[78] Boeing's strong wish to have the 737 Max planes certified under the terms of existing 737 NG certification led them to even seek and receive approval from the Federal Aviation Authority to remove all mention of MCAS from flight manuals - https://airwaysmag.com/industry/boeing/boeing-faa-cultures-to-blame-for-737-max-disasters/

[79] Lion Air Flight 610 on 29 October 2018 and Ethiopian Airlines Flight 302 on 10 March 2019.

[80] The wife of Lion Air Flight 610's Captain provided evidence at trial that her husband was actually trained by Boeing pilot trainers in America prior to receiving type approval and flying any 737 Max planes - thus discrediting Boeing's American-centric safety claim.

planes completely and sales of the plane had ceased, before Boeing acknowledged their mistakes and began to investigate approaches to resolving the issue with their autonomous system. Blaming the human occupant is a common theme seen even with semi-autonomous and self-driving motor vehicles. The Autopilot system in Tesla vehicles has in at least eleven cases[81] failed to identify parked emergency vehicles and as a result have killed occupants by driving into these stationary fire engines. Tesla CEO Elon Musk has repeatedly taken to social media and the mainstream media to deny issues with his company's autopilot self-driving system[82], misdirect by discussing how few injuries one driver received in these crashes[83], and even suggesting the occupant may be lying when they say autopilot was enabled[84].

Systems exist for scanning vehicle registration plates[85]. Motorway and intersection fixed cameras also record the registration plate and a timestamp for every car that passes under them whether the vehicle is speeding or has run a red light, or not[86]. In some cases fixed camera information is used to calculate a vehicle's average speed between multiple fixed points[87], in others it is simply recorded as part of the process to measure traffic flow. These technologies enable law enforcement and local authorities to verify in near real-time that vehicles are appropriately registered, and whether taxes or fees and periodic maintenance checks for the vehicle are *up to date*. The same infrastructure could be used to issue and verify the vehicle's digital driver's license as well. Where a manufacturer has met all requirements, passed all tests and had the vehicle type certified, a digital license could be issued over-the-air to allow the vehicle to *self-drive*. Without this digital license, the vehicle would either not travel, or might require the human driver to assume control and manually drive to their destination. Potential issues with this approach might include where the system data is in error and either erroneously revokes the vehicle's DDL, or alternatively issues a DDL to a vehicle that should not have one. Also, it could strand unlicensed occupants[88] relying on the autonomous functionality if the vehicle ceases to self-drive partway into the journey as a result of either DDL revocation or because the traffic network has become inaccessible[89]. However, this technology also creates potential privacy, security and abuse of data issues[90]. While some law enforcement agencies and municipalities claim the data is not retained, it has been shown that this data is being retained and that these government organisations are collecting sufficient data to build vast histories from people's journey data[91]. Leaving these privacy and security issues aside, it is possible that these systems could support a future where vehicles have electronic registration rather than the current process that requires a human owner to ensure registration fees are paid and that maintenance checks have been performed. When a vehicle is observed by a camera, the system could perform the checks and, where necessary, issue an over-the-air alert to the vehicle where there is a lapse that requires attention.

---

81 https://arstechnica.com/cars/2021/08/us-investigates-autopilot-after-11-teslas-crashed-into-emergency-vehicles/ and https://www.theguardian.com/technology/2021/aug/16/teslas-autopilot-us-investigation-crashes-emergency-vehicles

82 https://innovation-village.com/police-demand-tesla-crash-data-as-musk-denies-autopilot-use/

83 https://twitter.com/elonmusk/status/996132429772410882

84 https://twitter.com/elonmusk/status/1384254194975010826?lang=en

85 Known as Automated Number Plate Readers (ANPR). These systems can recognise using OCR and record number plates in live camera feeds - whether the camera is attached to police cars, the entrances of parking structures or at service station forecourts. Police ANPR can automatically query vehicle registry databases to return details of the registered owner or keeper and current regulatory status of the vehicle.

86 https://www.schneier.com/blog/archives/2008/07/speed_cameras_r.html and https://www.csoonline.com/article/3562472/lessons-learned-from-the-anpr-data-leak-that-shook-britain.html

87 Average Speed Limit detection.

88 Including drivers who only have the proposed *Certificate to Operate an Autonomous Vehicle* discussed in the previous paragraph. Such a certificate is more like a certificate to oversee the safe operation of the machine and based on the researcher's descriptions, would not convey the right or ability to actually subsume control from the AI.

89 While we do not discuss it here, there also exists the issue that some might seek to jailbreak the vehicle and override such functionality. While this would not update the central systems that provide DDLs, it could allow the vehicle to override the lockout and operate. Further, Cory Doctorow and others have pointed out that simply passing laws to say jailbreaking is illegal is not a solution to this problem. He who controls the software code controls the vehicle, and people exist everywhere who will want to unlock and potentially tinker with or seek to understand how their device operates.

90 https://melbactivistlegal.org.au/2020/09/24/automated-number-plate-recognition-anpr-surveillance-during-covid-19/

## 4.3 The Presumption of Safety

A recent *barriers, benefits and facilitating factors* literature review found improvement to vehicle safety realised by a significant reduction in traffic accidents was the most often cited benefit that adoption of autonomous cars would bring[92]. While some believe autonomous cars will be almost entirely accident free[93], others set the bar much lower - believing *safer than the average human driver* to be a sufficient benchmark (Nees et al, 2019). Such autonomous safety studies are most often found in the domains of psychology and ethics, respectively investigating safety as a question of public perception using questionnaires (Liu, Wang & Vincent, 2020; Nees et al, 2019; Penmetsa et al, 2019), or as an ethical dilemma using trolley problems (Bonnefon et al, 2019; Hubner & White, 2018; Nyholm & Smids, 2019). Both domains, and approaches, completely avoid dealing with the limitations either inherent in or introduced to big data *machine learning* technologies[94]. Similarly, while many who author research on autonomous cars use language claiming or promoting improved safety (and other benefits), these claims remain untested, unrealised and the product of journalist and industry-funded think-tank imaginations of what development and adoption of smart technology-enabled cars should be like[95].

Another point often overlooked is that autonomous car safety may be inextricably linked to another issue: *security* of the vehicular hardware, communications networks and software that will support autonomous vehicle operation.

## 4.4 The Three Fallacies of Security

During the last decade tech companies have made many grand claims regarding the escalating technological measures being integrated into their products, that we are told will ensure or enhance the security of our devices and personal information[96]. These include: (a) use of

---

[91] https://www.autoexpress.co.uk/news/106295/massive-anpr-camera-data-breach-reveals-millions-private-journeys

[92] See note 81 - McLachlan et al, 2022.

> "That SDC would improve road safety was a strong claim in 63% of works, with authors believing improved safety would manifest in a reduction of traffic accidents (49%) and that fewer accidents would translate to fewer deaths (29%)."

[93] Schneble, C. O., & Shaw, D. M. (2021). Driver's views on driverless vehicles: Public perspectives on defining and using autonomous cars. *Transportation research interdisciplinary perspectives*, *11*, 100446.

> Interviewees in the study reported in this paper discussed that intelligent driverless self-driving cars would be integrated, more reliable, react faster and *per se quite accident free*.

[94] Cory Doctorow argues that the focus on trolley problems is being used to avoid the issues of computer and software flaws and government-mandated back doors that will all weaken security and create known vulnerabilities in autonomous vehicle systems. See: https://www.theguardian.com/technology/2015/dec/23/the-problem-with-self-driving-cars-who-controls-the-code

[95] See note 71. McLachlan et al.

[96] Early examples include Apple's 2010 announcement of the first TouchID fingerprint sensors in the iPhone 5s that they stated was a "smarter way to secure a user's iPhone". Early TouchID systems suffered from an insufficient number of 'points' representing key locations in the user's fingerprint. This resulted in examples of young children and even cats being able to unlock the 'secure' devices. See: https://appleinsider.com/articles/13/09/10/apple-announces-touch-id-fingerprint-scanner-for-iphone-5s and https://www.quora.com/How-does-my-20-month-old-daughter-unlock-my-iPhone-which-is-fingerprint-protected and https://www.dailymail.co.uk/sciencetech/article-2425504/The-video-shows-CAT-unlock-new-iPhone-5S-paw-used-instead-fingerprint.html

passcodes, fingerprint and facial recognition into smartphones and laptops; (b) integration of security and encryption technology directly into web servers[97], software applications, network appliances, file storage and even motherboards[98]; and (c) ubiquitous virtual private network (VPN) services. To date, almost none of these measures arrived without issue, and the added security they delivered tended to be temporary or fleeting[99]. This example is representative of our first fallacy of security - *that a device, software application or smart technology can ever be totally secure*.

The constant ongoing battle between Apple and IOS jailbreakers provides a perfect example of the first fallacy - that creating the next seemingly *most secure* version of computer hardware or software might close one *known* security hole - but in doing so adds more software code and potentially creates new vulnerabilities that result either from the new code itself, or the way the new code interacts with existing code in the device. Recent reports showing IOS jailbreaks now working on Apple's *MacBook Pro* laptops demonstrate how cross-pollination by the vendor of hardware or operating system code between platforms can result in known vulnerabilities being transported into what are ostensibly different platforms[100].

There has been a constant war between those who wish to keep their personal information private - and outsiders who wish to access it. No one would be surprised to find hackers have tried to access their personal files. However, it is ironic that law enforcement and governments have used public service announcements (PSA) to advise us to protect our files and online banking sessions using end-to-end encryption[101], while also (a) purchasing software hacking tools that they install and operate on government-owned servers in order to circumvent that same encryption and access our devices and data remotely[102]; and passing laws requiring tech

[97] HTTPS first used Secure Sockets Layer (SSL) and, when vulnerabilities were discovered in SSL, was more recently replaced with Transport Layer Security (TLS) to encrypt data in transit to and from web servers and the web browser.

[98] Unified Extensive Firmware Interface (UEFI) replaced Basic Input/Output System (BIOS) and increased the security of core hardware in computer systems by enabling such services as Secure Boot to restrict those applications that can be used at boot time and help computers resist attacks and infections by malware. The Trusted Platform Module (TPM) provides a range of hardware-based security-related functions including to generate, store and limit the use and changing of secure encryption keys, improve authentication through the use of an integrated RSA key, and through the taking and storage of security measurements of the system's hardware and software.

[99] As seen in note 75, TouchID came with issues and continues to have software bugs today (https://forums.macrumors.com/threads/huge-issue-with-touchid-on-2021-macbook-pro.2326769/ and https://thehackernews.com/2020/08/apple-touchid-sign-in.html). The more advanced criminal or law enforcement can even use forensic tools to identify and then make a mask of your fingerprint lifted directly from the TouchID button (https://medium.com/@3raxton/how-touch-id-a-security-flaw-cc-apple-35b131de369a). While Apple's T2 Security chip, similar to the TPM chip used in Windows devices, was supposed to provide hardware-level security, researchers and jailbreakers found a vulnerability that not only allowed unlocking of T2-locked devices, but also one which exposed the encryption keys of the hardware to hackers and law enforcement (https://www.wired.com/story/apple-t2-chip-unfixable-flaw-jailbreak-mac/). It is possible that this was by design in order to provide the legislated 'backdoor' required in those countries, like Australia, New Zealand, Russia and India, that have laws requiring a backdoor in encrypted hardware and software. And issues with VPNs have included that VPN companies have used weak and unsafe encryption that was easily reversible (HolaVPN in Dec 2018; Hotspot Shield, PureVPN and Zenmate VPN also in 2018); VPN companies who promise security and anonymity and not to keep logs who then admit to providing complex and complete logs to law enforcement (NordVPN in Jan 2022); and VPN company servers that are quietly taken over by law enforcement (DoubleVPN in June 2021) and used temporarily as a honey trap (VPNLab in Jan 2022).

[100] https://www.wired.com/story/apple-t2-chip-unfixable-flaw-jailbreak-mac/

[101] For example: https://www.esafety.gov.au/key-issues/how-to/protect-personal-information and https://www.ncsc.gov.uk/collection/10-steps/data-security and https://ico.org.uk/for-organisations/guide-to-data-protection/guide-to-the-general-data-protection-regulation-gdpr/security/encryption/ and https://www.consumer.ftc.gov/articles/how-safely-use-public-wi-fi-networks

[102] It has long been known that government and law enforcement teams in many countries have purchased and installed servers running Italian hacking group Hacking Team's Galileo and Da Vinci software that allows remote exploit injection (hacking), control, and full data access of a target individual's mobile devices. Devices attacked in

companies to provide back doors into encrypted software[103]. More recently, some governments have signed multinational public safety policies that are largely about making it easier for the surveillance services and law enforcement to have unfettered access to our personal information[104]. These governments have funded misleading public advertising campaigns described as *disingenuous scaremongering*[105] and *creepy*[106] that, in contradiction to their own PSAs on personal data security, are intended to induce anxiety in the public about any use of end-to-end encryption. They claim common software applications like Facebook's Messenger (now known as Meta) and Signal using end-to-end encryption leave your children vulnerable to child abuse and other crimes, put everyone at risk, and prevent law enforcement from detecting crime[107] while never telling you that the same encryption is what keeps your online banking transactions secure and free from prying eyes. They never tell you why the government want unrestricted access to every citizen's communications and personal information. Nor do they tell you that the absence of end-to-end encryption will significantly weaken your security and data protection when accessing your online banking, applying for loans or credit cards, or doing any of the other perfectly legitimate things that end-to-end encryption is used to secure. This issue is representative of the second fallacy of security - *that governments and law enforcement will act in our digital best interests and maintain our right to digital privacy and security*.

In discussing autonomous car security issues, Cory Doctorow recently asked how long it will be before government and law enforcement demand backdoors integrated into these vehicles so that they can remotely force the vehicle off the road when initiating a traffic stop[108]. He says that legislatively-mandated provision of technology backdoors for "lawful interception" simply weakens overall security and are a gift to criminals who will also access these backdoors to take control of your self-driving vehicle.

Finally, digital integration of systems, services and functionality has become popular. Rather than having to access two or more bank websites to view your different accounts, Bank A will allow you to enter the details of your accounts at Bank B and you can then authorise for those accounts to also be accessible from within Bank B's website. This is an example of inter-bank digital integration. As noted in section 4.1, governments are pushing citizens to use digital apps that bring together a number of State services and functions. This is an example of government and State-services digital integration. An example that may be even more ubiquitous is that of the Apple and Android "wallet" Apps. These apps allow you to electronically collect your credit and debit cards, digital identification, digital tickets and other artefacts into a single application. *Inter alia*, a single app in your smartphone becomes your payment mechanism; digital driver's license; COVID passport; and bus, train or airline ticket. However, what if your smartphone is stolen? For example, your ex-partner breaks into your car, takes it[109], and knows

---

this way can be remotely activated so that not only can the device's location be immediately identified without accessing a cell provider's network, but also to secretly take photos with the camera or turn on microphones and listen to private conversations. https://www.wired.com/2014/06/remote-control-system-phone-surveillance/

[103] https://www.theverge.com/2020/10/12/21513212/backdoor-encryption-access-us-canada-australia-new-zealand-uk-india-japan and https://fee.org/articles/australia-s-unprecedented-encryption-law-is-a-threat-to-global-privacy/

[104] https://www.justice.gov/opa/pr/international-statement-end-end-encryption-and-public-safety#_ftnref1

[105] https://www.independent.co.uk/tech/encryption-government-campaign-home-office-b1995605.html

[106] https://www.eff.org/deeplinks/2022/01/uk-paid-724000-creepy-campaign-convince-people-encryption-bad-it-wont-work

[107] See note 83.

[108] https://www.theguardian.com/technology/2015/dec/23/the-problem-with-self-driving-cars-who-controls-the-code

[109] A Tasmanian mechanic used Jaguar Land Rover's InControl app to unlock and steal his ex-partner's smartphone from her locked SUV. https://www.whichcar.com.au/car-news/man-found-guilty-stalking-with-car-app

or easily guesses your unlock code. They can now use your digital wallet to make expensive purchases and, due to digital integration, the fraud prevention text messages your credit card provider sends that ask you to answer Yes or No to whether you are making the purchase, come up on the screen of the same device. They respond to the bank's service in the affirmative and are able to complete the transactions. Apps in your phone might let them open the digital door lock on your house, steal your car, access your email and social media, or order and pay for an expensive Uber ride to across town using your account. This issue is representative of the third fallacy of security - *that digital integration is nett beneficial and safe.*

As more key services and functionality are digitised and integrated into a single App or device, greater is the risk that one security issue will expose all your digital assets. Digital integration is nothing more than a modern example of the idiom: *putting all your eggs in one basket*.

## 5. Summary and Conclusion

The push for digitisation of State-issued identification like Driver's licenses needs further consideration. Presently, there are insufficient benefits to justify the risks and costs involved in a complete move to a ubiquitous and total digital public infrastructure. Further, the privacy issues and potential for State, corporate and criminal exploitation of the aggregate datasets of individual's personal information that nation states will build behind digital ID systems whose creation citizens of many countries have fought hard against have not, as yet, been adequately contemplated or addressed. Given the events of the last two years, now does not seem to be the right time to blindly move forward with legislation for these systems.

It has been said that fewer or greater numbers of driver's licenses may be issued in future. However, when we extrapolate based on comparisons to other transport domains that have high levels of automation like aviation, the future for driver's licenses may not be so *up in the air*. Introduction of autonomous systems in aviation frequently leads to pilot retraining, and while modern commercial jets are generally sufficiently capable to take off, fly to their destination and land with little in the way of human intervention, pilots are still required and every pilot must hold a current license with ratings for the types of plane they command. Contemporary in-service training for commercial pilots has shifted focus from extended practicing of general flight skills like navigation and radio use. Pilots now spend many hours each year in simulators training to building knowledge, experience and thus muscle memory for how to respond when the airplane's extensive interconnected collection of semi- and fully-autonomous fly-by-wire systems go awry. While the plane itself is actually *flying* for much of the time during most flights today, the pilot's license still attests to their ability to assume full control and complete the flight where it becomes necessary. We contend a similar future may exist for driver training and licensure.

# Appendix A: Test and Type Evolution - Per Country

**Canada (CA):** While the GDL concept was first proposed in Canada, and Quebec had an informal GDL from 1977, GDL were only officially introduced in 1994 (Dionne & Laberge-Nadeau, 2012). In 1998 British Columbia introduced their GDL and required learner drivers to display an L (learner) plate and, once they had passed their driving test and for two years thereafter, an N (novice) plate - effectively the same as most other country's P (provisional) plates (Moneo, 2010).

**Germany (DE):** The first German practical driving test resulted from a Prussian ministerial decree in 1903[110]. The *Reich Law on Motor Vehicle Traffic* 1909 introduced uniform traffic rules, a general issue driver's license, and regulated training services and driver testing[111]. Written multiple choice tests were introduced in 1956[112]. In 1986 a probationary license was introduced[113].

**United Kingdom (UK):** Like NZ, early drivers licenses in the UK were issued simply as a tool to identify vehicles and their drivers (Finnerty, 2015). Practical driving tests were introduced in 1935. Written theory tests were introduced in 1997 and computerised and presented on a touchscreen from 2000[114]. Introduction of practical driving tests also created the need to buy a provisional license and the same regulation introduced L (learner) plates for drivers who were yet to pass the test and receive a full license (DVSA, 2019). Provisional licenses were introduced for heavy goods vehicle (HGV) drivers in 1937 (DVSA, 2019). Computerised theory testing were extended to include Hazard Perception (HP) tests in 2002 (Senserrick & Williams, 2015). Unlike most other countries based on the English Common Law system (Australia, New Zealand and Canada), the UK abandoned introduction of a GDL (Chillingsworth, 2020), preferring to enhance existing approaches with a new structured driving lesson scheme and additional restrictions and curfews for novice drivers (Geoff, 2021).

**New Zealand (NZ):** While cars had to be registered and licensed from 1905 (Pawson, 2014), annual licensing of the driver was not required until 1925 (Pawson, 2014). Practical driver testing was introduced in 1931 (Begg & Stephenson, 2003). The first written and oral driver theory tests were introduced in 1965 (Price, 2012) with computerised theory tests trialled in 1990 in Dunedin (Price, 2012). Computerised learner driving tests were legislatively introduced for every new driver in 2009[115]. NZ has chosen not to adopt Hazard Perception testing. In 1966 a probationary (learner) license system was introduced and new license holders were classed as *probationary* drivers for their first 2years - they had to wear an L plate and were limited to a maximum speed of 50mph (Begg & Stephenson, 2003). The 1966 probationary system was found to have had no effect on the NZ crash rate and was repealed in 1971 (Begg & Stephenson, 2003). Prior to the GDL a 15 year old could receive a full unrestricted license on completion of a standard driving test comprising of: (i) eyesight and hearing tests; (ii) 25 written questions; (iii) five oral questions; and (iv) a practical driving test (Begg & Stephenson, 2003). The GDL was introduced in 1987 (Bates et al, 2014) and as at 2016 NZ was still operating the traditional three-stage GDL (Learner → Provisional → Full/Open License).

[110] https://www.autobild.de/artikel/100-jahre-fuehrerscheinpruefung-43515.html
[111] https://second.wiki/wiki/geschichte_des_fc3bchrerscheins
[112] https://www.autobild.de/artikel/100-jahre-fuehrerscheinpruefung-43515.html
[113] https://second.wiki/wiki/geschichte_des_fc3bchrerscheins
[114] https://blog.passmefast.co.uk/driving-law/driving-licence-history/
[115] https://www.odt.co.nz/news/national/hardest-road-code-questions-revealed

**United States of America (USA):** New York State was the first in 1910 to issue a driver's license on paper that came with an attached photograph (Gruffyydd, 2014). In 1908 Rhode Island introduced USA's first practical driving tests (Gruffyydd, 2014) and from 1913 New Jersey required written and oral tests prior to issuing a driving license[116,117]. New York State established an early Learner's Permit in 1925 (Gruffyydd, 2014). While most States have had some rudimentary elements of a GDL from 1921 (Gruffyydd, 2014), the first true multi-stage GDL were introduced in 1996 (Dionne & Laberge-Nadeau, 2012; Williams, 1997). Automated computerised testing systems were implemented in Florida from 2001 (FLHSMV, undated). However, like NZ, USA have not adopted Hazard Perception tests.

**Australia (AUS):** Commenced practical driver testing in 1911. While Australian drivers had to pass an oral test from 1938, written tests only commenced in 1982[118]. Hazard perception tests were introduced first in Victoria in 1996, and gradually in other states (Senserrick & Williams, 2015). AUS had a learners license from the 1940's and a rudimentary form of legislated GDL with Learner and Provisional licenses in operation from 1966 in NSW (*Motor Traffic Regulations Act* 1965) that required a period on a provisional license with several restrictions for all 'new' drivers (Faulks & Irwin, 2009) and some previous drivers who have had their licenses cancelled[119] - including the need to display a "P" plate on their vehicle[120]. The first proper GDL was introduced in 1990 in Victoria (Dionne & Laberge-Nadeau, 2012) and the current version was updated in 2000 (Faulks & Irwin, 2009). While AUS began with the traditional three-level graduated system (Learner → Provisional → Full/Open license), by 2016 three states (ACT, SA and TAS) have each since added a Pre-Learner phase (Pre-Learner → Learner...), TAS has added two phases of Learner license (Pre-Learner → Learner1 → Learner2...), and all States and Territories except NT had adopted two-phases of provisional license (...Provisional1 → Provisional2 → Full/Open License) (Scott-Parker, 2016). This means that in some of the highest-regulated states (ACT and TAS) the new driver may undergo five phases over more than five years before graduating onto a full/open license (Pre-Learner → Learner1 → Learner2 → Provisional1 → Provisional2 → Full/Open License) (Scott-Parker, 2016).

**France (FR):** France introduced the world's first legislation providing for a mandatory practical driving test in 1893 (Carter, 2003). Pearson VUE was awarded the contract to deliver the new French computer-based driving test in 2016[121], which continues to be recognised as one of the more difficult computer-based driver theory tests[122].

---

116 https://magazine.northeast.aaa.com/daily/life/cars-trucks/auto-history/the-history-of-the-drivers-license/
117 https://911drivingschool.com/the-drivers-license-then-and-now/
118 https://www.rms.nsw.gov.au/documents/about/environment/protecting-heritage/oral-history-program/vehicle-regulation-driver-testing-and-licensing-summary-report.pdf
119 NSW Department of Motor Transport - Commissioner for Motor Transport Annual Report 1965-66, https://www.opengov.nsw.gov.au/download/16169
120 Ibid.
121 https://www.pearsonvue.co.uk/About-Pearson-VUE/Press-Room/2016/Pearson-VUE-announces-Pointcode.aspx
122 https://www.thelocal.fr/20160504/french-stumped-by-new-driving-test/